\documentclass[letter,11pt]{article}
\pdfoutput=1 

\usepackage{jheppub} 

\usepackage[LGR, T1]{fontenc} 
\usepackage{amsmath}

\usepackage{amsmath}
\usepackage{sourcesanspro}
\usepackage{mathptmx} 
\usepackage{booktabs} 
\usepackage{tikz}
\usepackage{feynmp-auto}
\usepackage{lineno}

\usepackage{placeins}
\usepackage{subcaption}

\notoc 

\usepackage[perpage]{footmisc}

\title{\huge\textbf{\\[2cm] Quantum Chromodynamics at the\\ Large Hadron Collider }}

\author{
Jesse Liu$^a$, on behalf of the ALICE, ATLAS, CMS, LHCb Collaborations\\ [0.2cm]
\emph{$^a$Department of Physics, New York University,\\ 726 Broadway, New York, NY 10003, USA}
}

\abstract{Open questions on the fundamental nature of the strong force endure and the Large Hadron Collider (LHC) is a once-in-a-generation laboratory elucidating its quantum origins. 
This document summarizes the plenary overview talk titled ``QCD Studies at the LHC'' presented at the Lepton Photon Symposium 2025. 
Selected results highlight recent experimental advances in Quantum Chromodynamics (QCD) at the LHC. 
This reviews the breadth of QCD and its cross-cutting synergies from a particle physics perspective in four themes: terascale precision tests, non-perturbative enigmas, mystery of confinement, and extreme cosmic-ray puzzles. 
}

\begin{document} 
\maketitle
\flushbottom

\newpage


\section{Introduction}
\FloatBarrier
\setcounter{page}{1}

Strong force dynamics are governed by the Lagrangian for Quantum Chromodynamics (QCD):
\begin{equation}
    \mathcal{L}_\text{QCD} = -\frac{1}{4}G_{\mu\nu}^a G^{\mu\nu}_a + \bar{q} (\mathrm{i}\gamma^\mu D_\mu -m_q)q + \frac{\theta}{32\pi^2}G_{\mu\nu}^a \tilde{G}^{\mu\nu}_a. 
\end{equation}
This is a Yang-Mills theory~\cite{PhysRev.96.191} whose quark $q, \bar{q}$ and gluon $G_{\mu\nu}$ fields are representations of the SU(3)$_\text{C}$ Lie group. 
Despite this theoretical simplicity, profound mysteries and open questions endure:
Who ordered three colors~\cite{Greenberg:1964pe,ALEPH:1992fwh,OPAL:2001klt}?
Why are there six colored fermions~\cite{Gell-Mann:1964ewy,E598:1974sol,SLAC-SP-017:1974ind,E288:1977xhf,CDF:1995wbb,D0:1995jca}?
Why is the strong coupling $\alpha_\text{QCD}$~\cite{Ahmadova:2024emn,CMS:2023fix} much larger than its electromagnetic counterpart $\alpha_\text{EM}$~\cite{Parker191,L3:2005tsb}?
What is the structure of its vacuum that gives the mysteriously unobserved topological $\theta$-term underpinning the strong charge-parity (CP) problem~\cite{Wilczek:1977pj,tHooft:1986ooh}? 
How are strongly-coupled theories tested~\cite{Muroya:2003qs,Bierlich:2022pfr}?

Confinement is arguably the defining hallmark of the strong force~\cite{Wilson:1974sk}, which endows the cosmos with nucleons rather than free-streaming quarks and gluons. 
It is therefore intertwined with the mystery of why nuclei, atoms, and life exist.
Studying QCD means probing the mysteries of our cosmological origins in the distant past~\cite{Wagoner:1966pv,Yang:1983gn}, high-energy cosmic rays~\cite{PierreAuger:2017pzq,TelescopeArray:2023sbd} and stellar extremes~\cite{LIGOScientific:2017vwq,LIGOScientific:2017ync} of today, and laboratory puzzles of why the neutron lacks an electric dipole~\cite{Abel:2020pzs}.

QCD is ubiquitous at hadron colliders.
The Large Hadron Collider (LHC) is no exception, unraveling the structure of matter and forces above TeV energies as a unique QCD laboratory. 
It is likely the last terascale collider in a lifetime. 
Theoretical models~\cite{Bierlich:2022pfr} help unravel the quantum structure of the enigmatic cascade from the terascale nine orders of magnitude down to confinement scale of hadrons (Figure~\ref{fig:eventdisplays} left).
This contribution highlights selected recent results from the LHC experiments.
Section~\ref{sec:precision} covers perturbative precision terascale tests before Section~\ref{sec:nonperturbative} moves to non-perturbative enigmas. Section~\ref{sec:confinement} discusses recent hadron results in the mystery of confinement then Section~\ref{sec:cosmics} summarizes QCD astroparticle synergies in extreme cosmic-ray puzzles before concluding.

\section{\label{sec:precision}Precision Terascale Tests}

\begin{figure}
    \centering
    \includegraphics[width=0.46\linewidth]{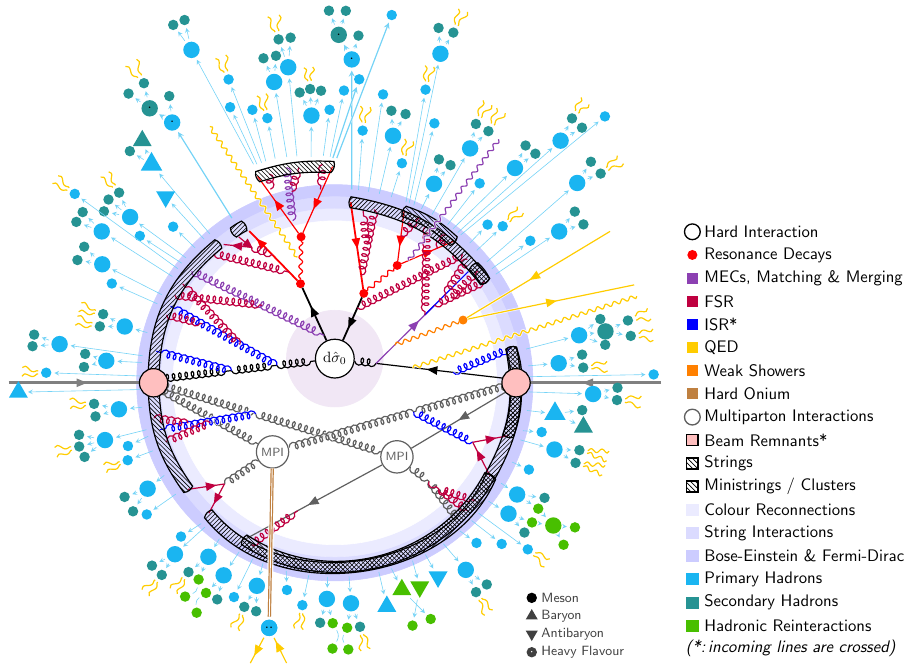}\quad 
    \includegraphics[width=0.5\linewidth]{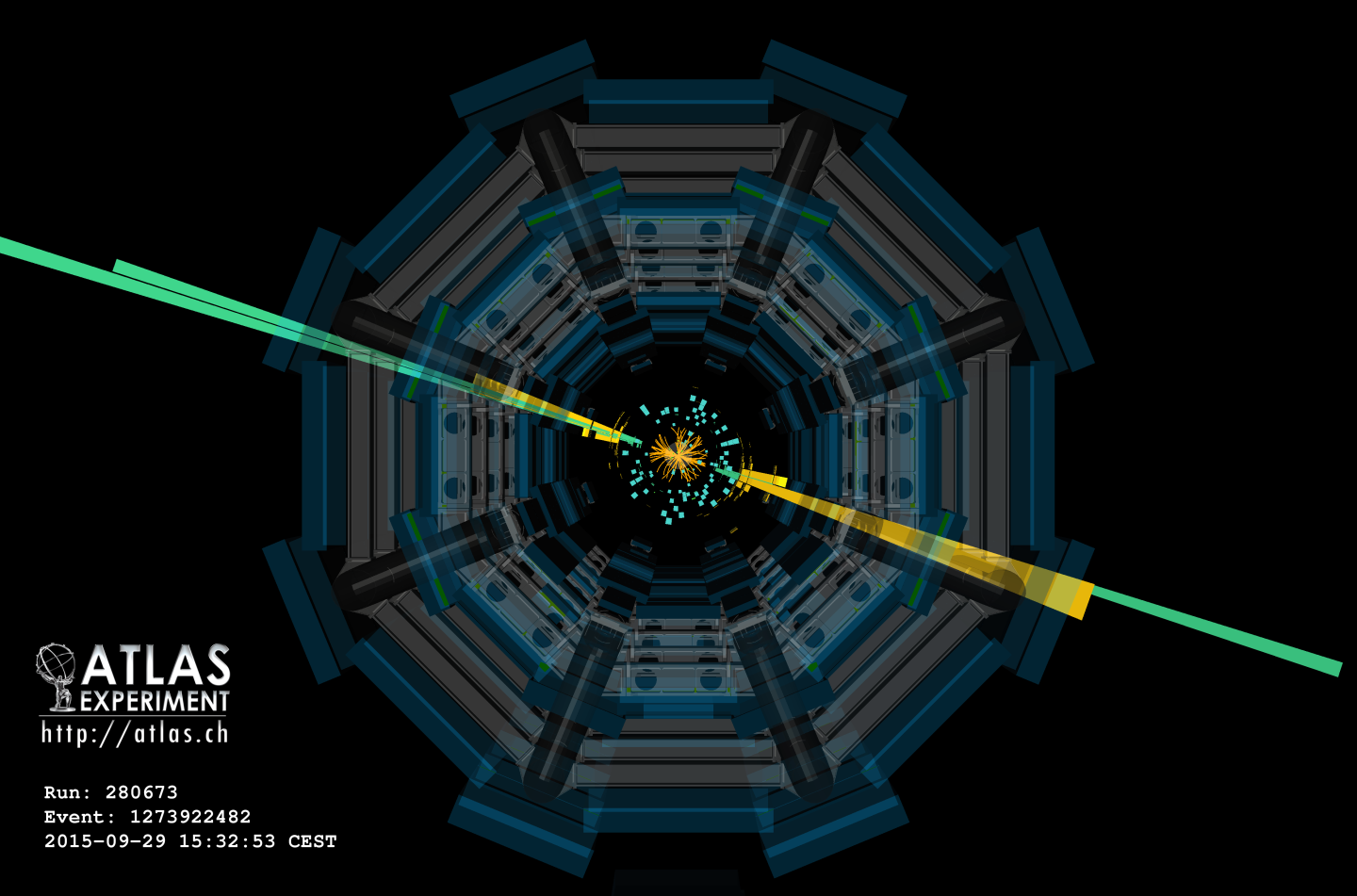}
    \caption{Phenomenological sketch of proton--proton collision~\cite{Bierlich:2022pfr} and dijet event display~\cite{Collaboration:2113239}.}
    \label{fig:eventdisplays}
\end{figure}

\begin{figure}
    \centering
    \includegraphics[width=0.50\linewidth]{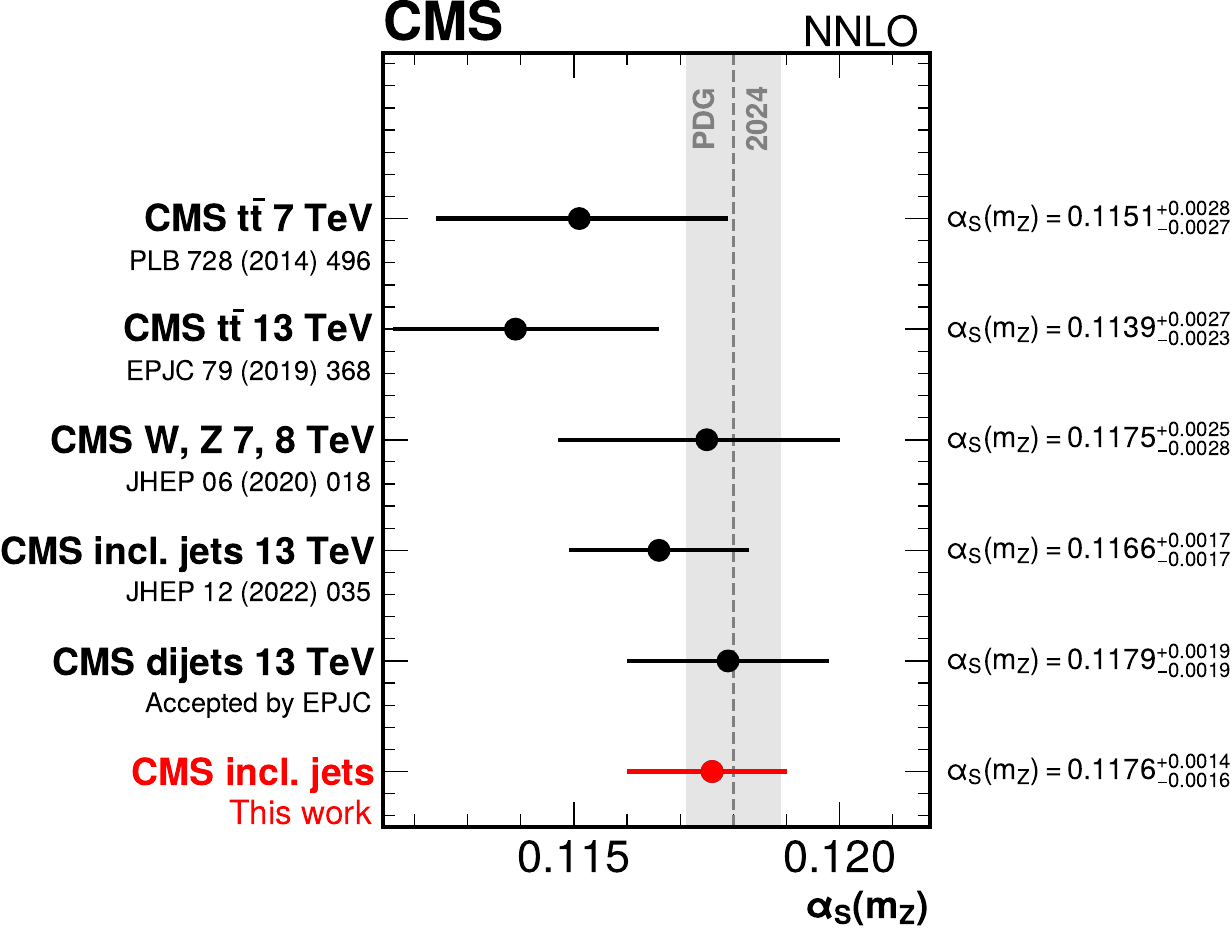}\quad
    \includegraphics[width=0.40\linewidth]{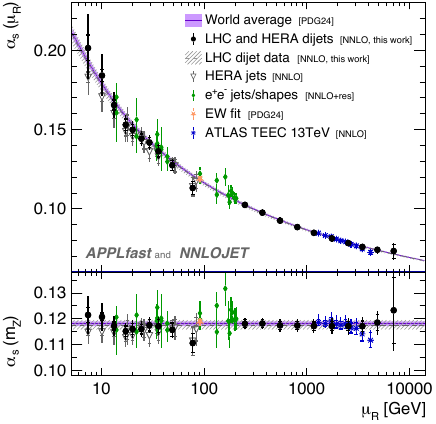}
    \caption{Recent measurements of the strong coupling $\alpha_\text{S}$ and its running with scale~\cite{CMS:2024trs,Ahmadova:2024emn}.}
    \label{fig:strong-coupling}
\end{figure}

The central miracle of QCD is that the strong coupling $\alpha_\text{QCD}$ becomes weak above the proton scale.
QCD becomes predictive, calculable, and testable to high precision via perturbation theory. 
Collimated sprays of hadrons called jets are the poster child of this miracle (Figure~\ref{fig:eventdisplays} right). 
CMS has published multi-differential dijet measurements out to nearly multi-TeV mass scales~\cite{CMS:2023fix} (Figure~\ref{fig:strong-coupling}).
Hadron colliders uniquely probe colored states' scattering as foundational tests of color flow in quark-gluon scattering.
These test difficult quark-gluon scattering amplitudes now calculable to next-to-next-leading order, where electroweak corrections grow significant at high masses.

Jets also probe the strong coupling deep into the terascale to test the foundational hallmark of QCD: anti-screening and asymptotic freedom at ultraviolet scales~\cite{Gross:1973id,Politzer:1973fx}. 
Recent CMS inclusive jet results enable the most precise measurement of $\alpha_\text{QCD}$ at $m_Z$~\cite{CMS:2024trs} (Figure~\ref{fig:strong-coupling} left).
Precise determination of the strong coupling to multi-TeV scales is interesting for probing any inflections from new physics~\cite{Ahmadova:2024emn}.
QCD is a renormalizable theory implying a consistent ultraviolet theory~\cite{tHooft:1972tcz}.
But importantly, asymptotic freedom is not inevitable from first principles. 
The QCD beta function $\beta(\alpha_\text{QCD})$ is a function of color $N_c$ and fermions $N_f$:
\begin{equation}
    \beta(\alpha_\text{QCD}) = -(11N_c - 2N_f)\frac{\alpha_\text{QCD}^2}{6\pi}. 
\end{equation}
Theoretically, asymptotic freedom only occurs if $N_f < 11N_c/2$ for the QCD beta function to remain negative if the number of colored fermions is 16 or less.
It remains a deep mystery why nature chooses the SU(N) group with $N_c=3$ with $N_f=6$, enabling infrared confinement and life to exist.

\begin{figure}
    \centering
    \begin{subfigure}{0.4\textwidth}
    \includegraphics[width=\linewidth]{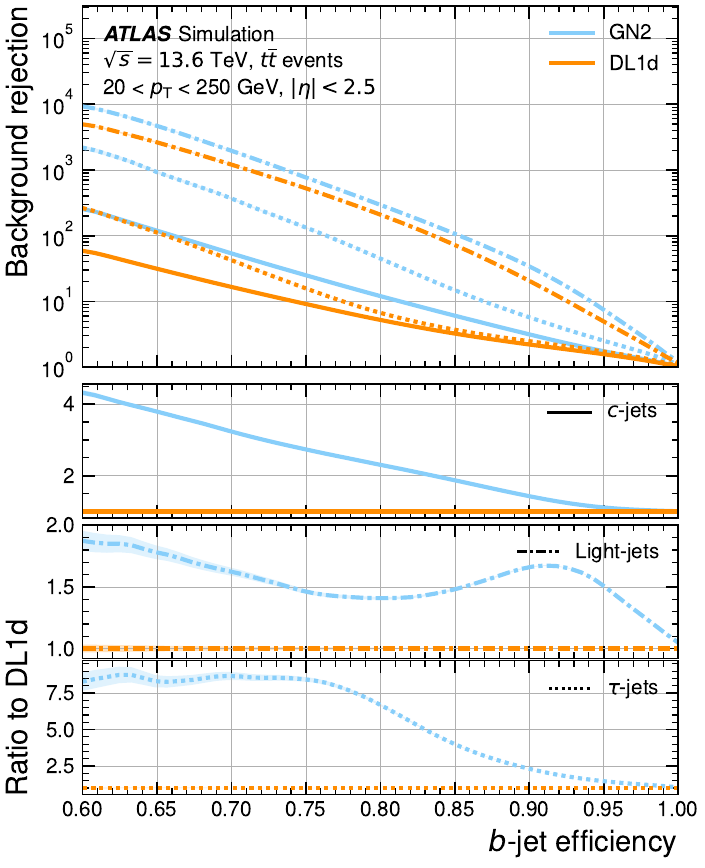}%
    \end{subfigure}%
    \begin{subfigure}{0.5\textwidth}
    \centering
    \includegraphics[width=0.8\linewidth]{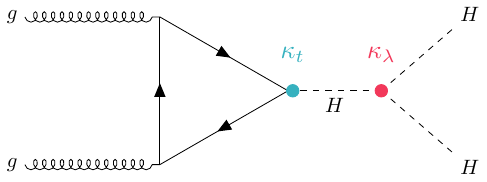}\\
    \includegraphics[width=\linewidth]{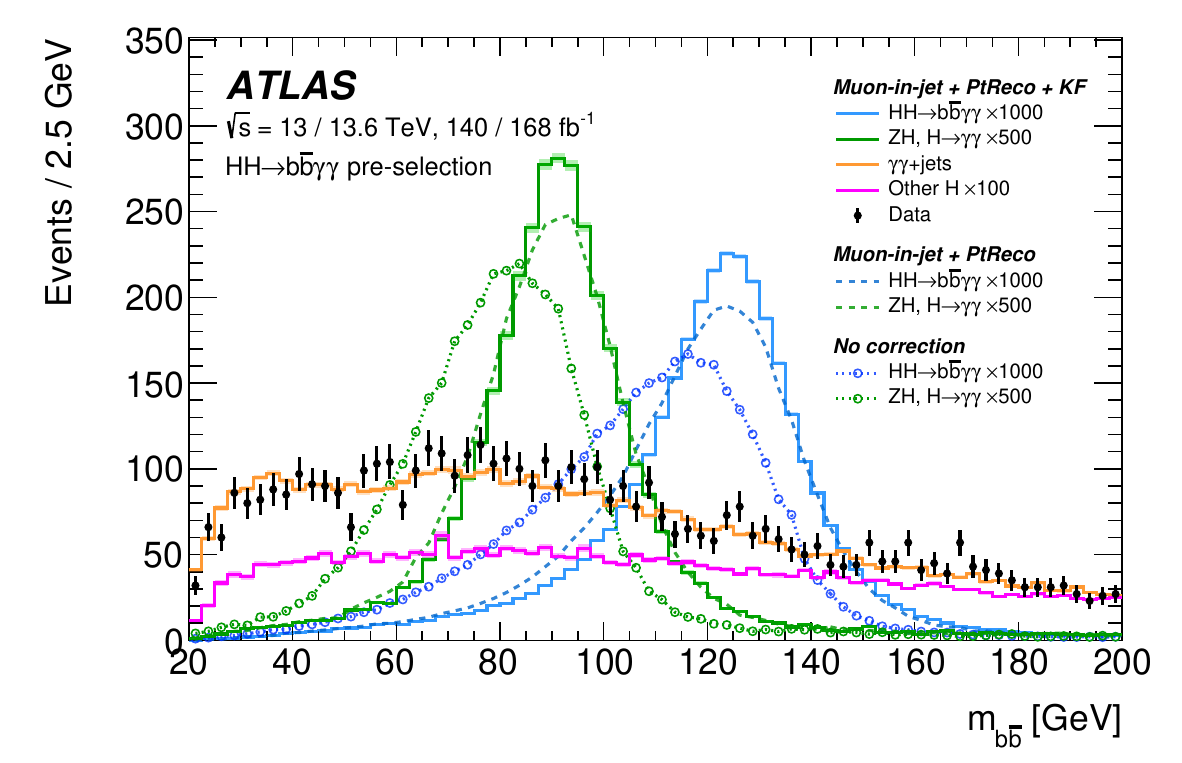}
    \end{subfigure}
    \caption{Jet flavor tagging advances~\cite{ATLAS:2025dkv} and applications to di-Higgs searches~\cite{ATLAS:2025hhd}.}
    \label{fig:flavor-tagging}
\end{figure}

\begin{figure}
    \centering
    \includegraphics[width=0.65\linewidth]{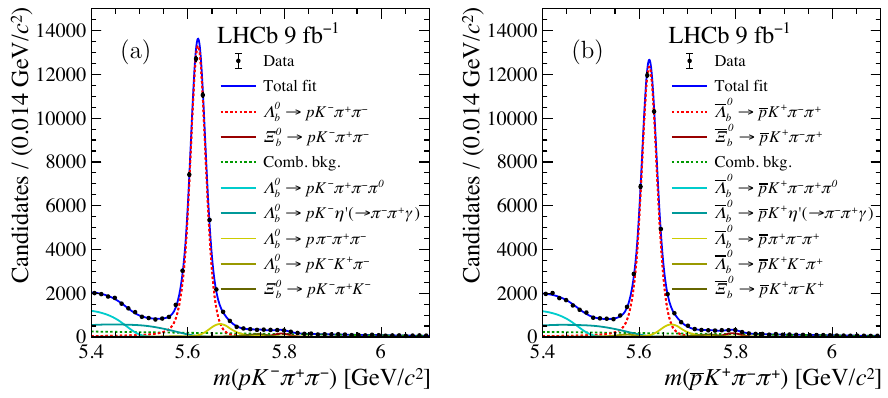}%
    \caption{Invariant masses of baryon and antibaryon modes for charge-parity analysis~\cite{LHCb:2025ray}.}
    \label{fig:lhcb-baryon-asymmetry}
\end{figure}

There is also striking progress in tagging jets originating from heavy-flavor quarks.
Flavor tagging is the machine learning testbed for the supervised classification problem. 
State-of-the-art taggers at ATLAS have moved from deep neural networks (DL1d)~\cite{ATLAS:2022qxm} for Run 2 to graph neural network transformers (GN2)~\cite{ATLAS:2025dkv} for leveraging full kinematic information.  
The GN2 tagger significantly improves rejection of charm, light and now even tau-lepton jets (Figure~\ref{fig:flavor-tagging} left).
These breakthroughs in reconstruction crucially accelerate the elucidation of electroweak symmetry breaking.
Di-Higgs is the central process enabling measurement of the Higgs self-coupling, which also motivates machine learning techniques~\cite{Amacker:2020bmn}.
ATLAS just released this 308~fb$^{-1}$ Run 2+3 combination in the $b\bar{b}\gamma\gamma$ channel~\cite{ATLAS:2025hhd} (Figure~\ref{fig:flavor-tagging} right).
Crucial to this progress are QCD advances: 
precision gluon parton distribution functions (PDFs), next-to-next-to-leading order predictions, graph network tagging, kinematic corrections improve the di-$b$-jet mass resolution to enhance sensitivity.

Heavy flavor also probes the mystery of baryon asymmetry in the universe.
Until recently, breaking of CP symmetry has only been observed in quark-antiquark mesons $q\bar{q}$.
However, the visible universe is dominated by three-quark states, namely baryons $qqq, \bar{q}\bar{q}\bar{q}$. 
LHCb has recreated this delicate effect experimentally for the first time~\cite{LHCb:2025ray}, observing $\Lambda_b$ baryons decaying at a slightly higher rate than its antibaryon counterpart (Figure~\ref{fig:lhcb-baryon-asymmetry}).
Observing this in the laboratory opens an important new path to probe CP symmetry breaking in the Standard Model and beyond.

Next is the basic question: how does the proton look when one zooms in with high definition?
It is far richer than the spherical blob of the 1950s~\cite{PhysRev.98.217} or static up-up-down cartoon from undergraduate textbooks~\cite{Gell-Mann:1964ewy}. 
At classic deep inelastic scattering energies, protons comprise quarks~\cite{PhysRevLett.23.935}.
At higher energies, it is dynamical and breaks scale invariance and can transform into an enigmatic cloud of gluons. 
A high definition proton means precision PDFs, crucial for precision tests of the Standard Model.
PDFs are key systematics from recent W boson mass measurements to the strong coupling~\cite{CMS:2024lrd}. 
There are dramatic differences in up/down valence quark and gluon PDF sets from a few to 50\% in these measurements of the strong coupling~\cite{CMS:2024trs}.

\begin{figure}
    \centering
    \includegraphics[width=0.33\linewidth]{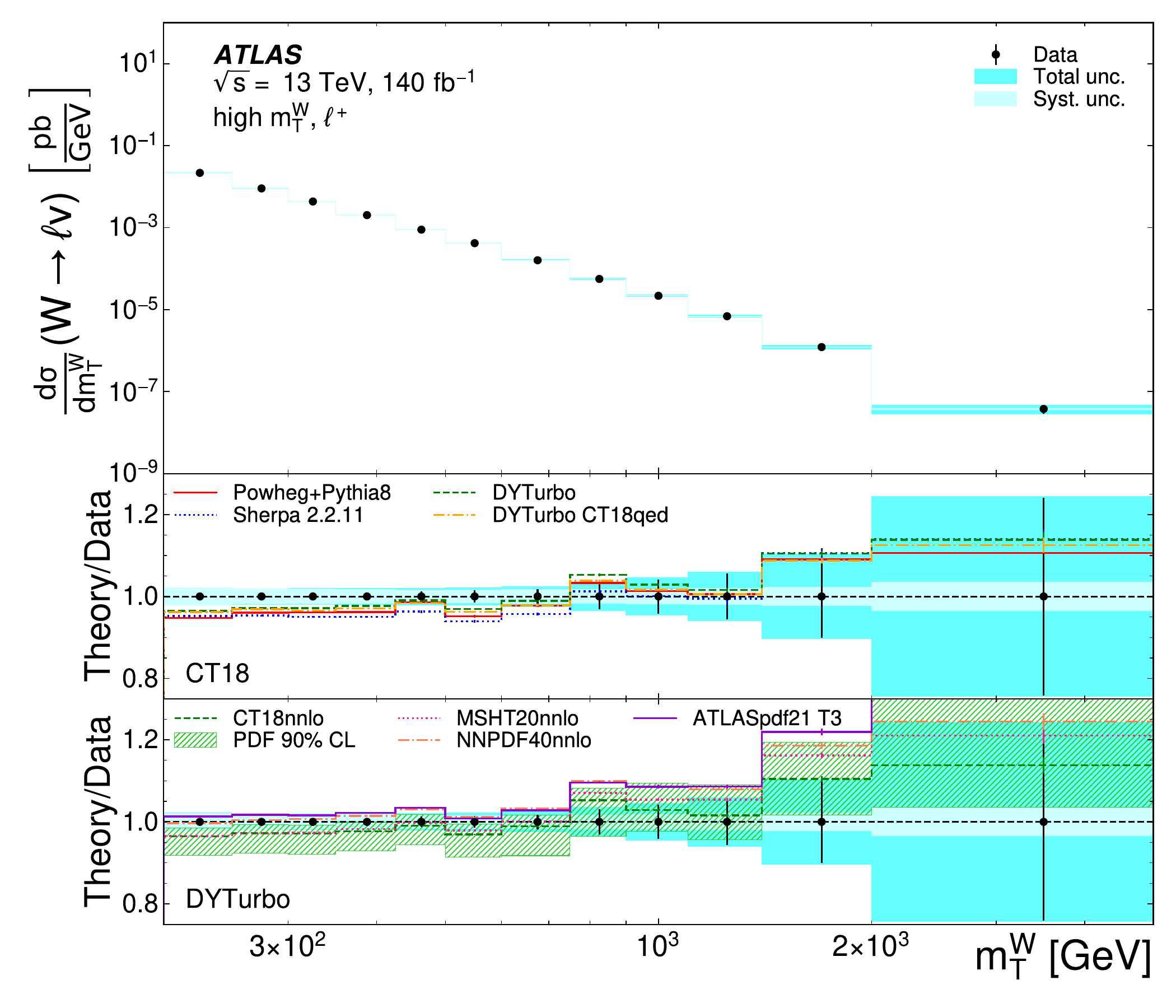}%
    \includegraphics[width=0.30\linewidth]{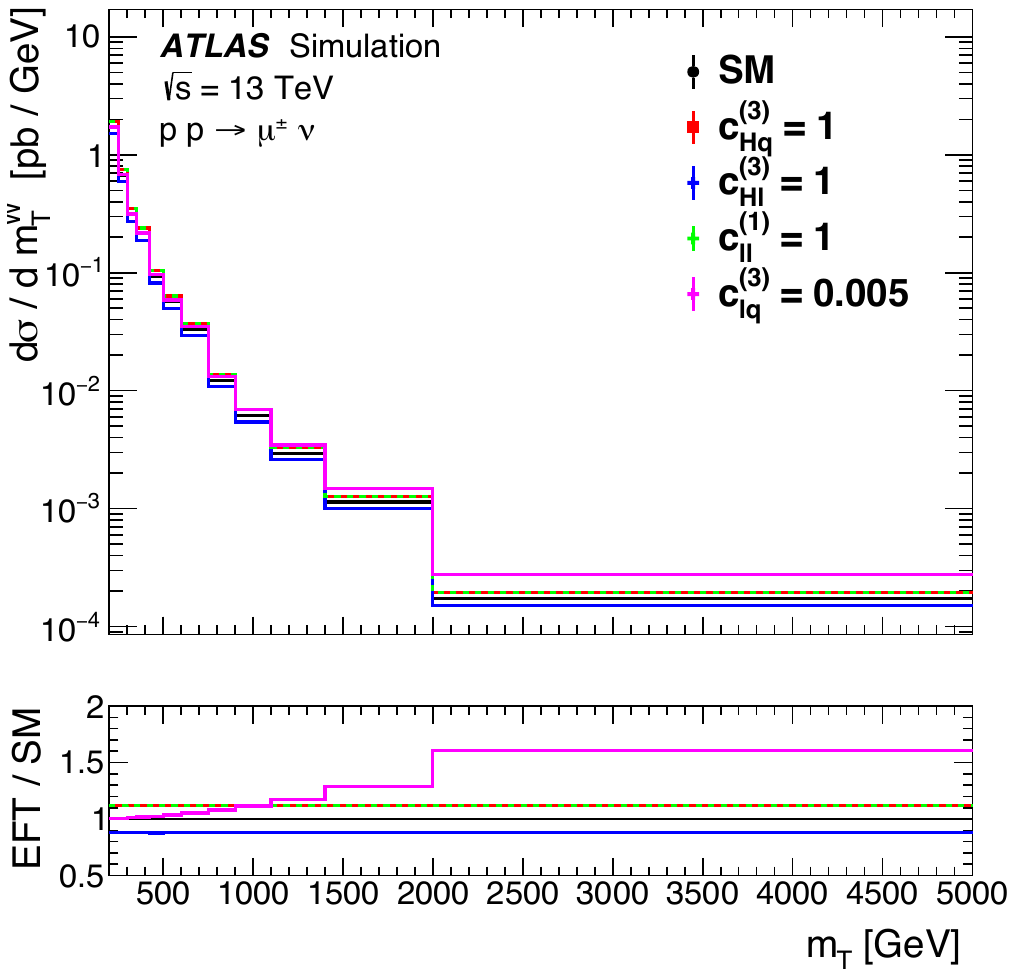}%
    \includegraphics[width=0.36\linewidth]{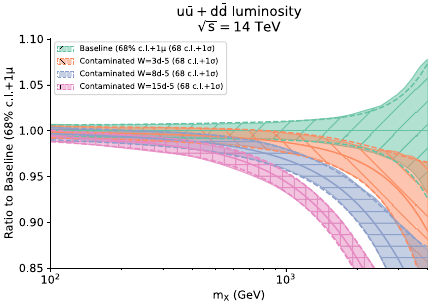}%
    \caption{(Left) High-mass $W$ measurements~\cite{ATLAS:2025hhn}, (center) new physics in tails~\cite{ATLAS:2025hhn}, (right) distorted parton luminosities~\cite{Hammou:2023heg}. }
    \label{fig:precision-tails-eft-pdf}
\end{figure}

It is therefore crucial to make more precise measurements that constrain PDFs.
ATLAS released a new measurement of boosted $W$ bosons~\cite{ATLAS:2025hhn} (Figure~\ref{fig:precision-tails-eft-pdf} left).
The cross-section of $W\to\ell\nu$ bosons differentially in transverse mass $m_\text{T}^W$ probes deep into TeV mass scales against state-of-the-art fixed order calculations and PDF sets. 
Measuring these tails is interesting for probing PDF uncertainties but also new physics contributions from Effective Field Theory (EFT) operators (Figure~\ref{fig:precision-tails-eft-pdf} center).
However, there are cautionary tales in tails. 
Theorists have recently raised provocative questions about whether PDFs could be fitting away new physics~\cite{Carrazza:2019sec,Hammou:2023heg}.
Injecting new physics into data can distort the up/down parton luminosities, impacting precision $WH$ measurements at the High-Luminosity LHC (Figure~\ref{fig:precision-tails-eft-pdf} right).
This conundrum in the precision EFT program requires disentangling tails from new physics simultaneous with probing phase spaces tha constrain PDF uncertainties.

\section{\label{sec:nonperturbative}Non-perturbative Enigmas}

\begin{figure}
    \centering
    \includegraphics[width=0.52\linewidth]{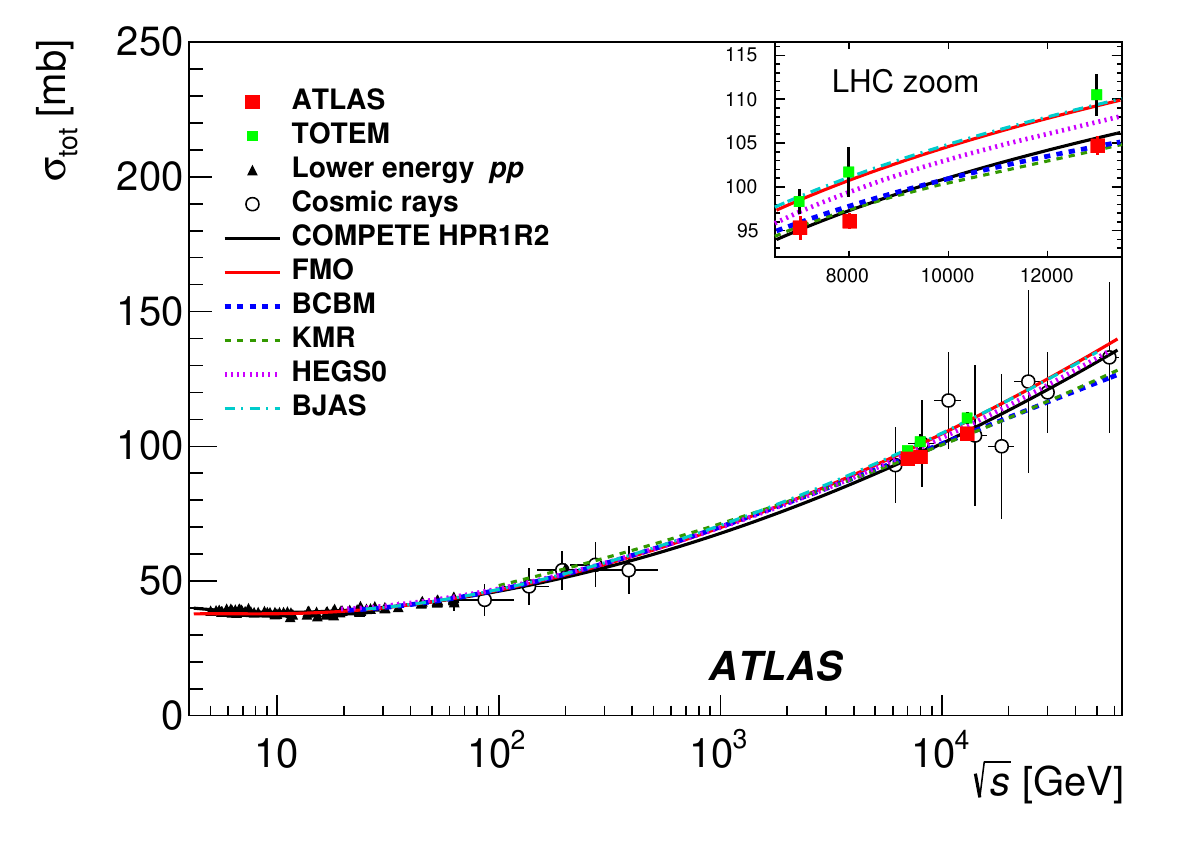}%
    \includegraphics[width=0.30\linewidth]{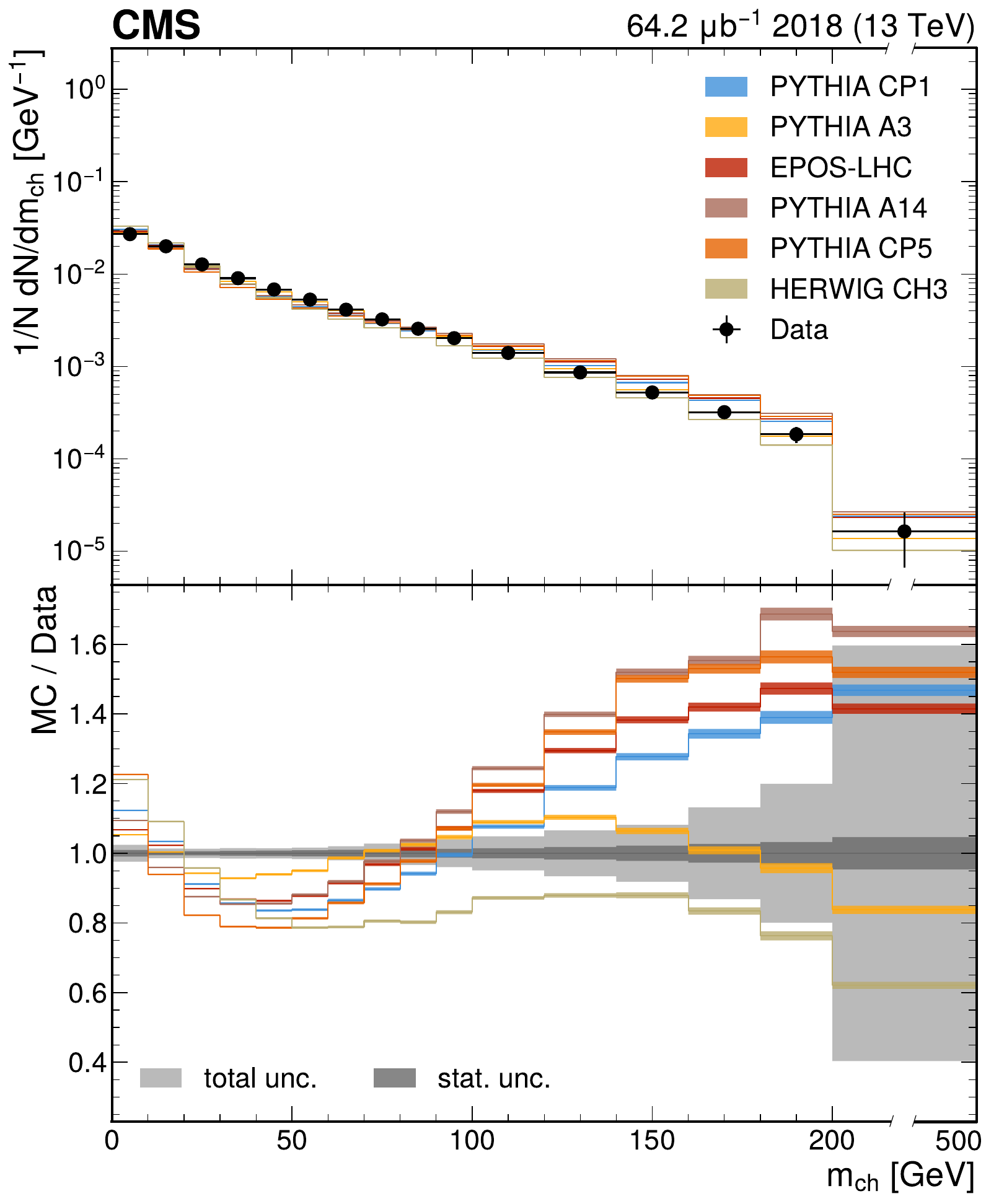}%
    \caption{Low-luminosity runs: total cross-section~\cite{D0:2020tig,ATLAS:2022mgx} and charged-particle distributions~\cite{CMS:2025sws}.}
    \label{fig:total-xsec}
\end{figure}

Next is the nightmare regime of $\alpha_\text{QCD}$ becoming strong.
Perturbation theory breaks down and QCD loses predictivity.
How can physics progress?
The answer is LHC data, which lift the terascale fog on strongly-coupled QCD. 
A cornerstone non-perturbative QCD measurement is the total proton--proton cross-section.
Low-luminosity data enable ATLAS and TOTEM~\cite{D0:2020tig,ATLAS:2022mgx} to probe deep into the terascale, extending 50 years of tradition since Intersecting Storage Ring (ISR) data~\cite{Amaldi:1973yv,Amendolia:1973yw} (Figure~\ref{fig:total-xsec} left).
There is a striking gap in laboratory data between ISR energies and 7 TeV, filled in only by cosmic-ray data.
Meanwhile, the lines are not first principles QCD calculations, but semi-empirical fits to phenomenological models.
These test foundational principles of locality, analyticity, and  unitarity captured in the Froissart bound~\cite{PhysRev.123.1053,Martin:1965jj}: $\sigma_\text{tot} \underset{s\to \infty}{\leq} C[\ln(s/s_0)]^2$. 
How does the cross-section grow? 
Perhaps medical advances may extend our lives another several decades until the next 100~TeV hadron collider to test if $\sigma_\text{tot}$ data continue rising or start decreasing beyond this bound?

Another non-perturbative QCD prediction is its topologically non-trivial vacuum with multiple degenerate vacua labeled by the Chern-Simons winding number~\cite{Callan:1976je,Jackiw:1976pf}.
Some theorists have braved calculating the QCD instanton tunneling rate, predicting isotropic sprays of particles at the LHC~\cite{Amoroso:2020zrz,Khoze:2021jkd}.
As scientists, it is prudent to ask if there exists any empirical evidence for this?
How can this be experimentally tested?
Minimum-bias collisions are a challenging background.
A new CMS measurement captures this challenge of 20 to 50\% spread between different generators in the charged-particle mass and sphericity~\cite{CMS:2025sws} (Figure~\ref{fig:total-xsec} right).
This remains a key challenge of instanton searches.

\begin{figure}
    \centering
    \includegraphics[width=0.35\linewidth]{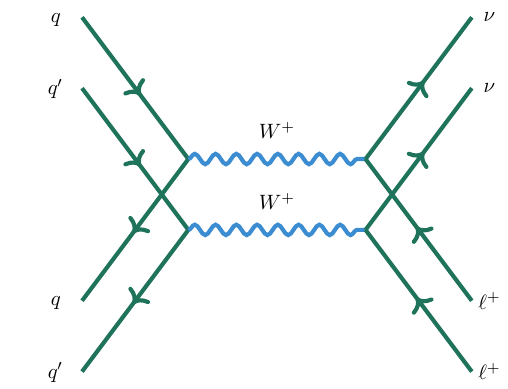}%
    \includegraphics[width=0.64\linewidth]{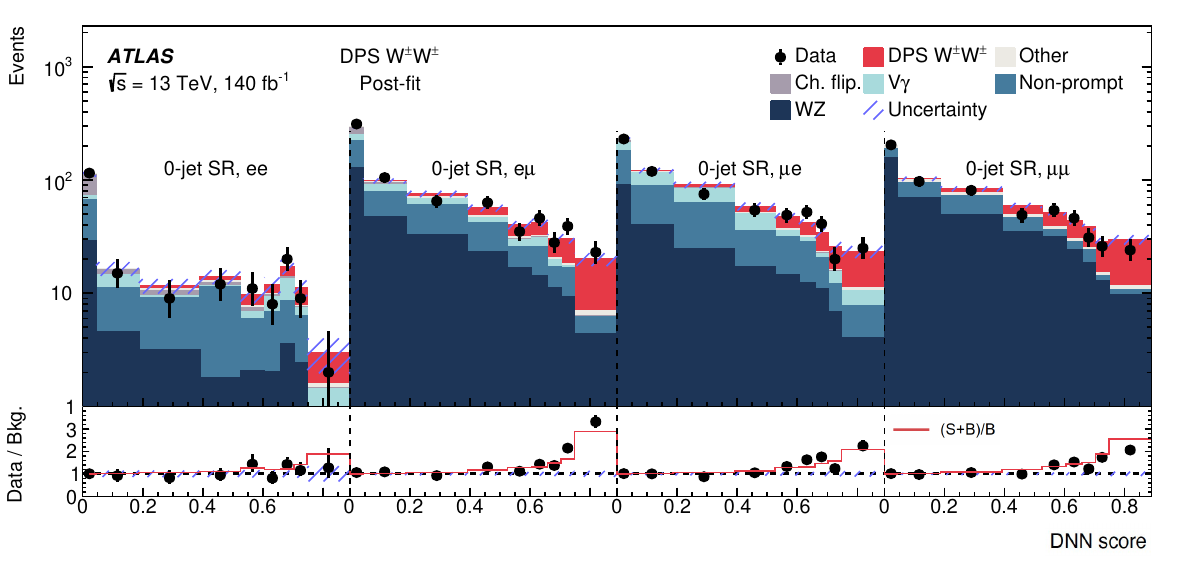}%
    \caption{Double-parton scattering in same-sign diboson production~\cite{ATLAS:2025bcb}.}
    \label{fig:ssww-dps}
\end{figure}

A more tractable non-perturbative QCD phenomenon is double parton scattering, long studied using clean $J/\psi\to\mu\mu$ data.
Usually just one quark per proton scattered, but sometimes two can scatter simultaneously. 
ATLAS observes two quarks in each proton interacting simultaneously to create two same-sign $W$ bosons~\cite{ATLAS:2025bcb} (Figure~\ref{fig:ssww-dps}), confirming the CMS observation~\cite{CMS:2022pio}.
This is important as a background to vector boson scattering $VV \to VV$ that probes electroweak symmetry breaking~\cite{ATLAS:2019cbr}. 

\begin{figure}
    \centering
    \includegraphics[width=0.25\linewidth]{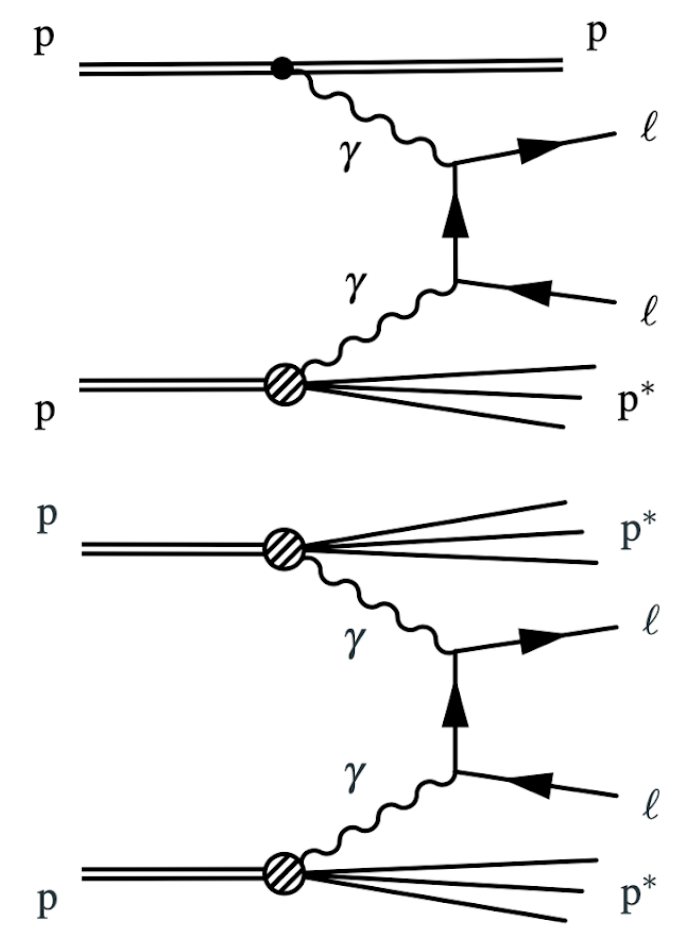}%
    \includegraphics[width=0.32\linewidth]{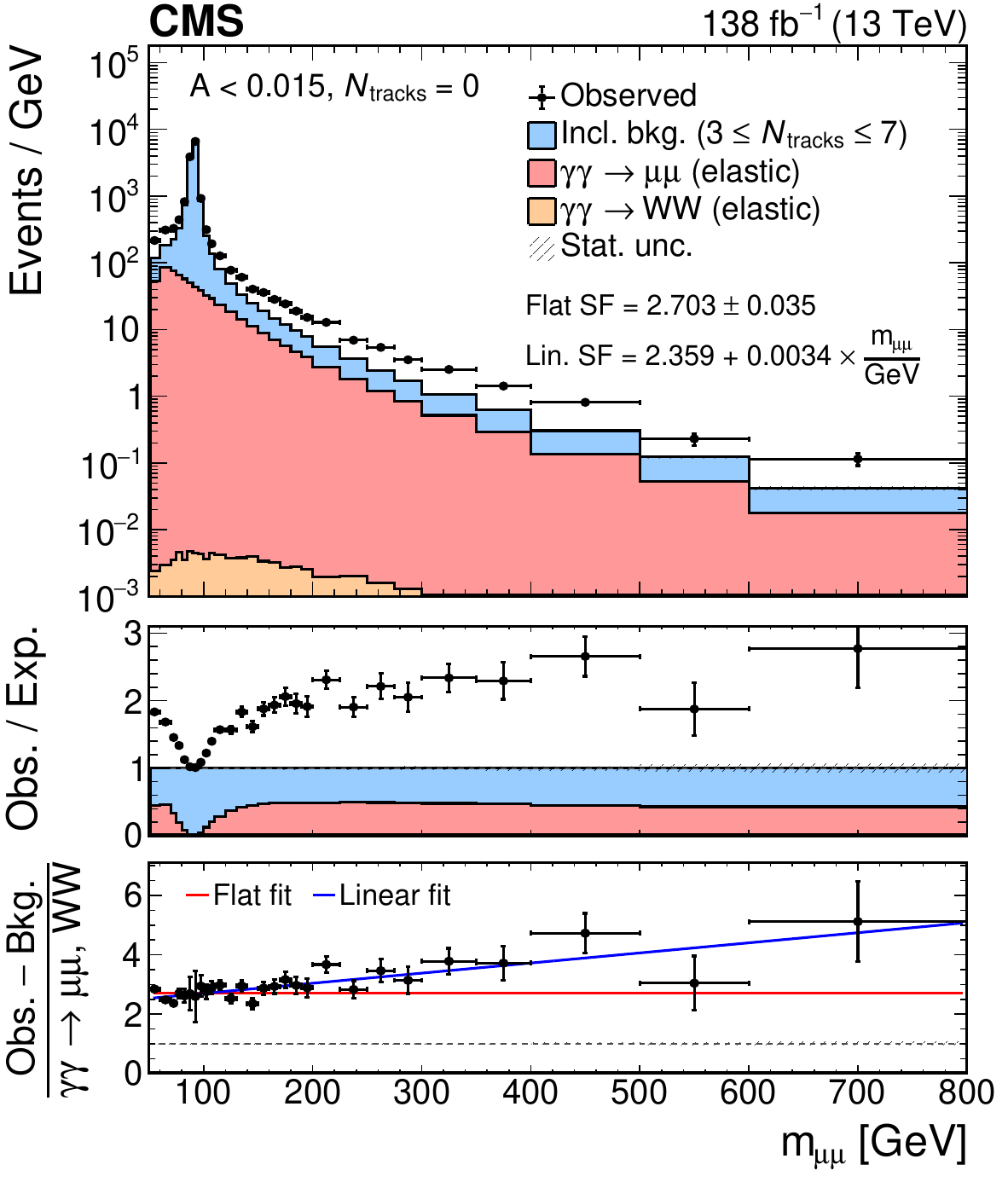}%
    \includegraphics[width=0.34\linewidth]{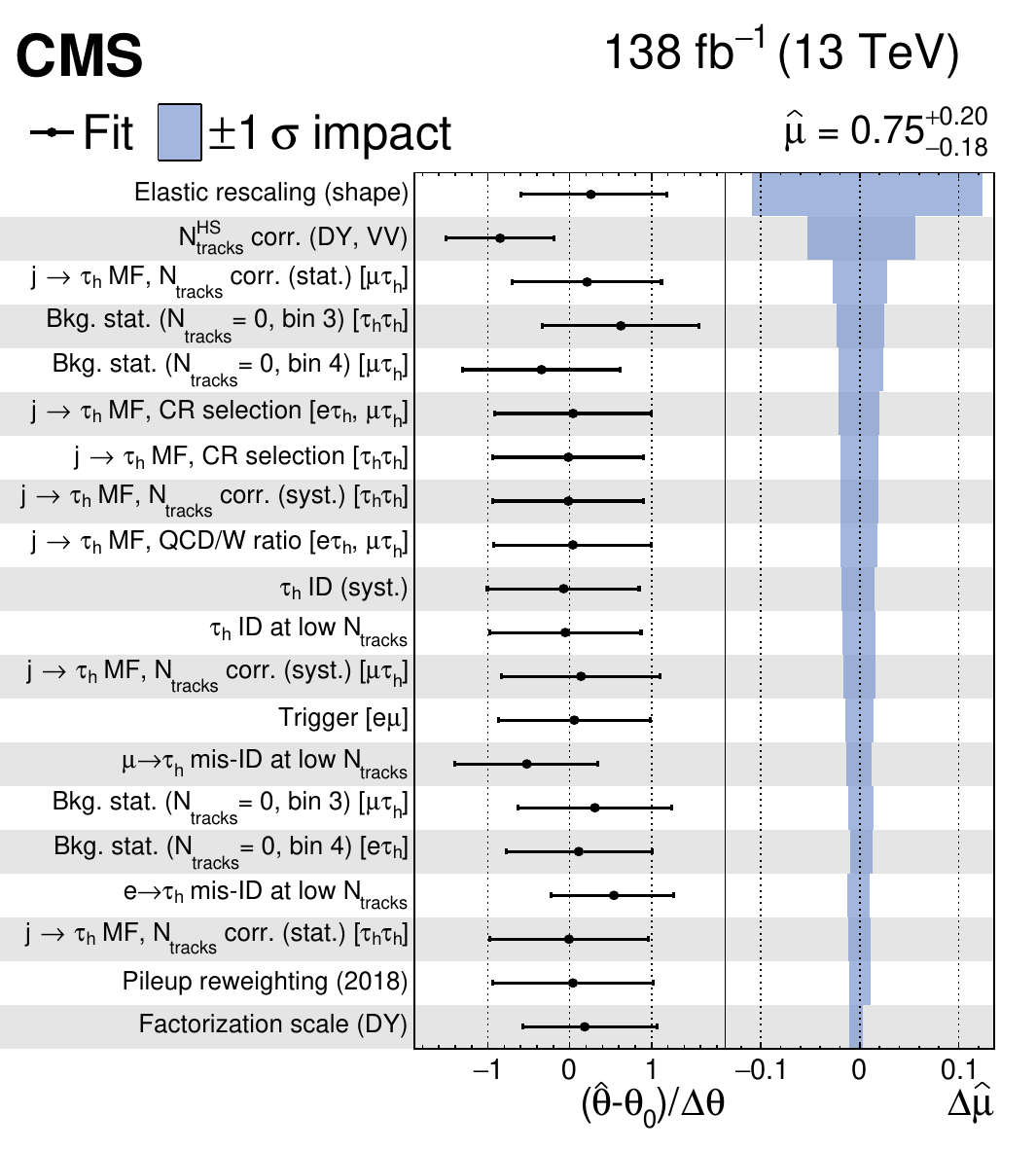}
    \caption{Diagrams of $\gamma\gamma\to \ell\ell/\tau\tau$, standard candle measurements of $\gamma\gamma\to \mu\mu$, and dominant systematics for the $\gamma\gamma\to\tau\tau$ measurement~\cite{CMS:2024skm}.}
    \label{fig:cms-photon-fusion}
\end{figure}

Recent advances turning the into LHC as photon collider are also limited by non-perturbative effects of proton breakup~\cite{Harland-Lang:2020veo,Shao:2022cly}.
These cannot be predicted from first principles and must be tuned to data.
Photon collisions are emerging as a promising novel direction in new physics searches~\cite{Beresford:2018pbt,Beresford:2019gww,Beresford:2024dsc,ATLAS:2022ryk}. 
Recently, the top two systematics in the CMS tau $g-2$ analysis via photon-induced tau-leptons $\gamma\gamma\to \tau\tau$~\cite{CMS:2024skm} (Figure~\ref{fig:cms-photon-fusion}) arise from proton-breakup uncertainties requiring standard candle calibration.
First, $\gamma\gamma\to ee/\mu\mu$ is used to constrain photon flux and proton dissociation effects~\cite{ATLAS:2020iwi,ATLAS:2020mve}.
Second, the underlying event from Drell-Yan is tuned to $Z\to ee/\mu\mu$ data to model hard-scatter proton breakup.

\begin{figure}
    \centering
    \includegraphics[width=0.25\linewidth]{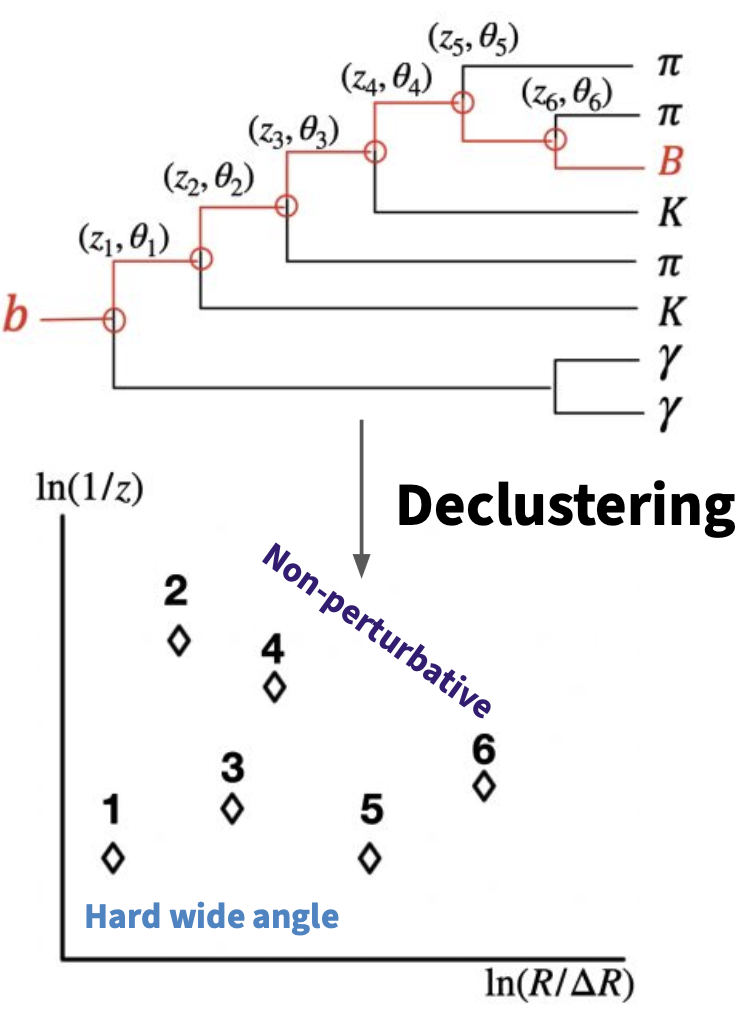}%
    \includegraphics[width=0.7\linewidth]{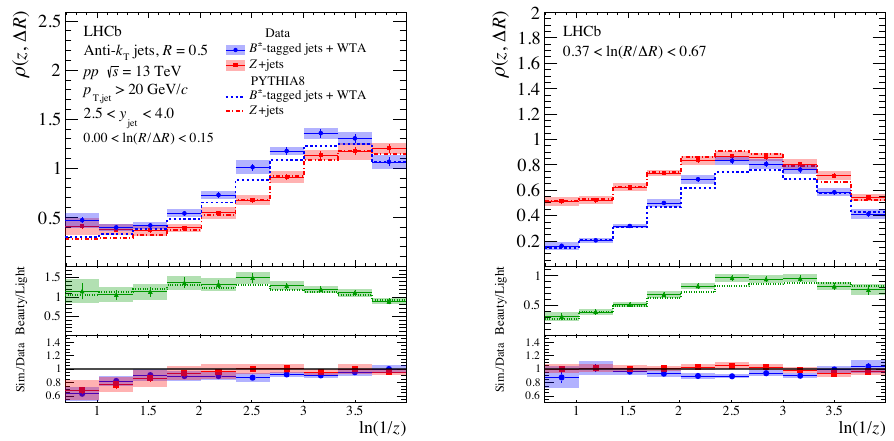}
    \caption{LHCb measurement of Lund Plane with flavor-tagged  and light jets~\cite{LHCb:2025mcq}.}
    \label{fig:lhcb-lundplane}
\end{figure}

\begin{figure}
    \centering
    \includegraphics[width=0.28\linewidth]{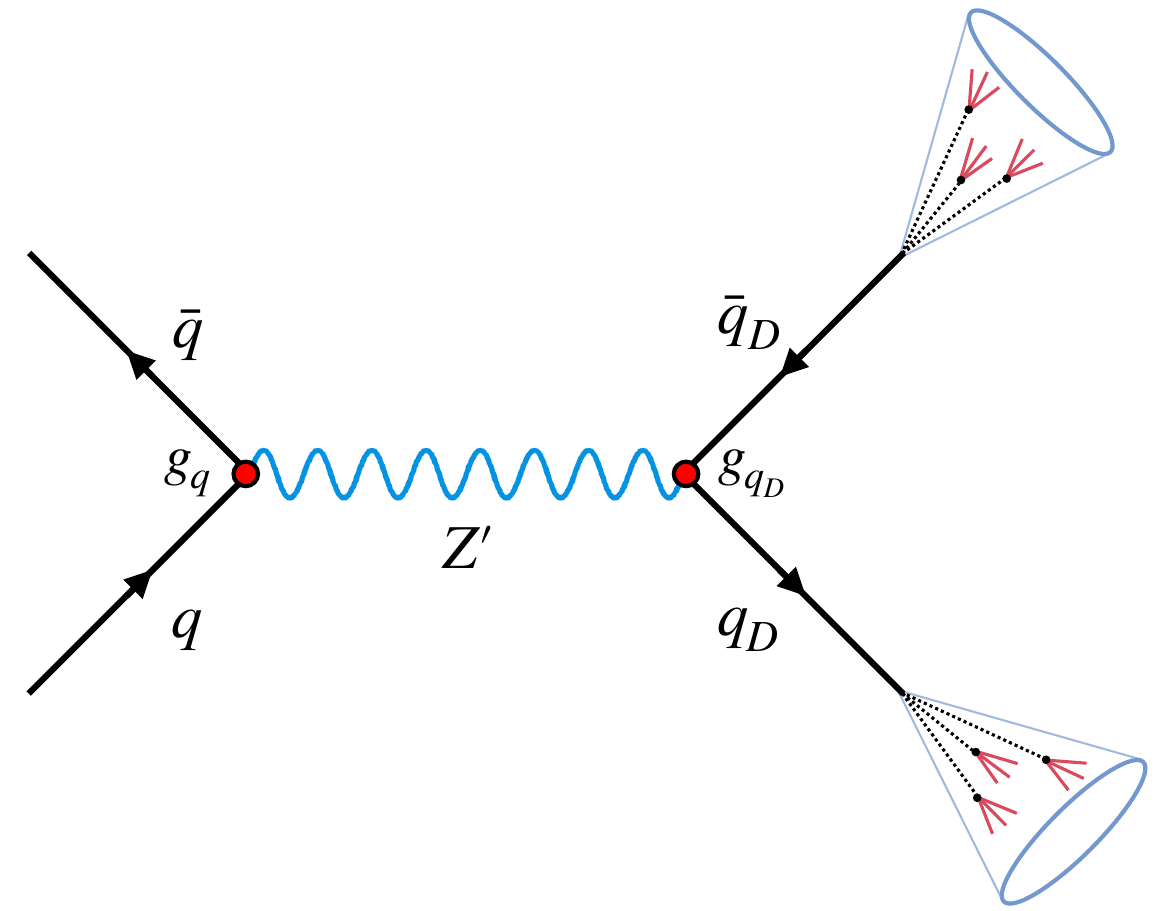}%
    \includegraphics[width=0.35\linewidth]{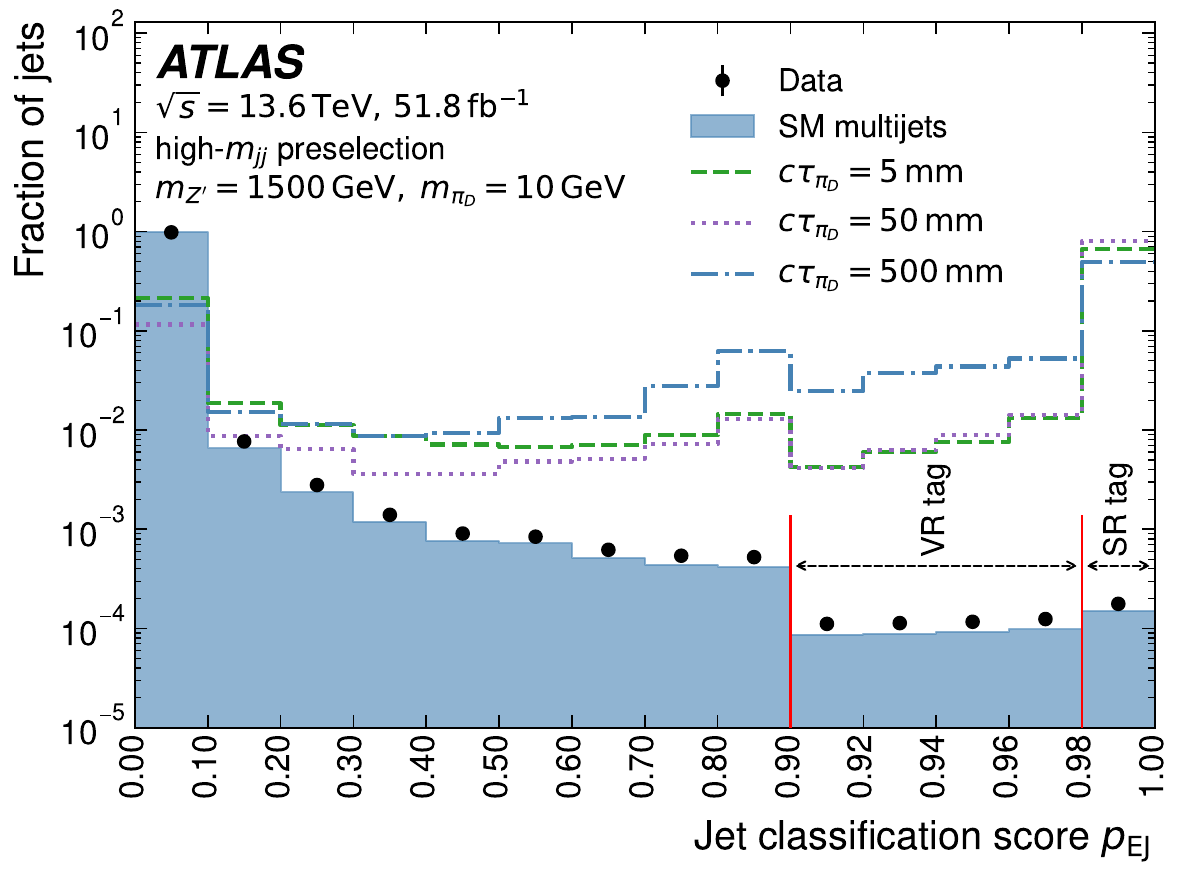}%
    \includegraphics[width=0.35\linewidth]{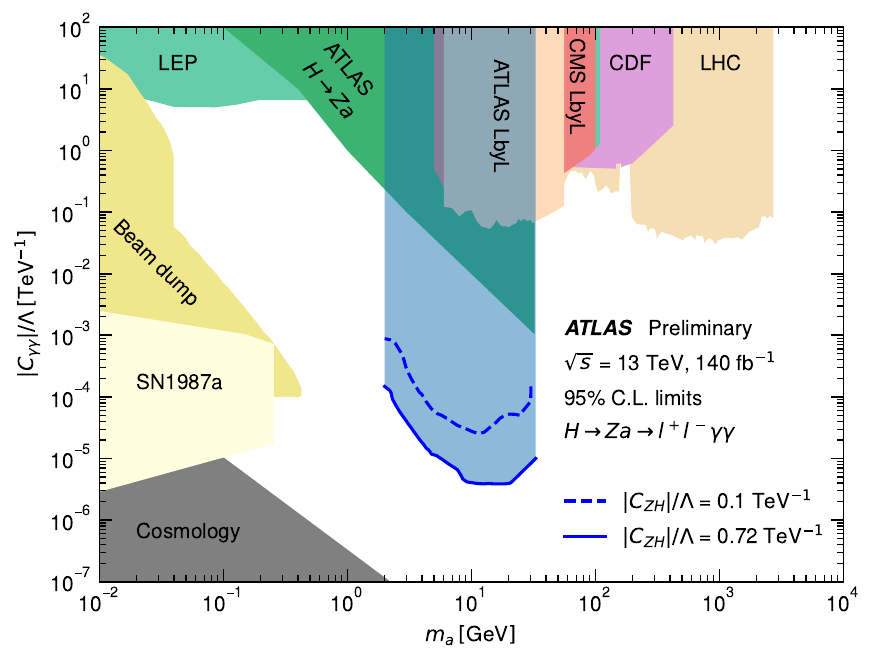}%
    \caption{Searches for dark QCD jets~\cite{ATLAS:2025bsz} and axions motivated by the strong CP problem~\cite{ATLAS:2023etl,ATL-PHYS-PUB-2025-007}.}
    \label{fig:dark-qcd}
\end{figure}

The parton shower captures the transition from perturbative parton to non-perturbative shower, where the Lund Plane systematically deciphers this shower history~\cite{Dreyer:2018nbf} (Figure~\ref{fig:lhcb-lundplane} left).
Extending light-jet only measurements~\cite{ATLAS:2024dua,CMS:2023lpp}, LHCb released the first measurement of the Lund Plane that directly compares bottom-quark tagged with light jets~\cite{LHCb:2025mcq} (Figure~\ref{fig:lhcb-lundplane} right).
Accurate parton-shower predictions are important because they are a prototype for dark QCD showers motivated by beyond-the-SM theories.
After all, if SM QCD showers are not well-modeled, how can modeling of dark QCD showers be trusted?
ATLAS released a new emerging jet search where there is a spray of displaced tracks (Figure~\ref{fig:dark-qcd} left).
The key innovation is deploying a dedicated Run 3 trigger adding a cut on low prompt track fraction to reduce $p_\text{T}$ cut from 500 to 200 GeV~\cite{ATLAS:2025bsz}, complementary other dark-sector searches~\cite{ATLAS:2025kuz,CMS:2024gxp}.
Meanwhile, the LHC is witnessing an axion renaissance motivated by the mysterious non-observation of a neutron electric dipole and strong CP problem~\cite{PhysRevLett.40.223,PhysRevLett.38.1440,Abel:2020pzs}.
Prompt and long-lived axions deepens diphoton sensitivity at the weak scale~\cite{ATLAS:2023etl,ATL-PHYS-PUB-2025-007} (Figure~\ref{fig:dark-qcd} right).
This complements sub-eV probes of axion-like particles~\cite{BREAD:2021tpx,GigaBREAD:2025lzq}.

\section{\label{sec:confinement}Mystery of Confinement}

\begin{figure}
    \centering
    \includegraphics[width=0.6\linewidth]{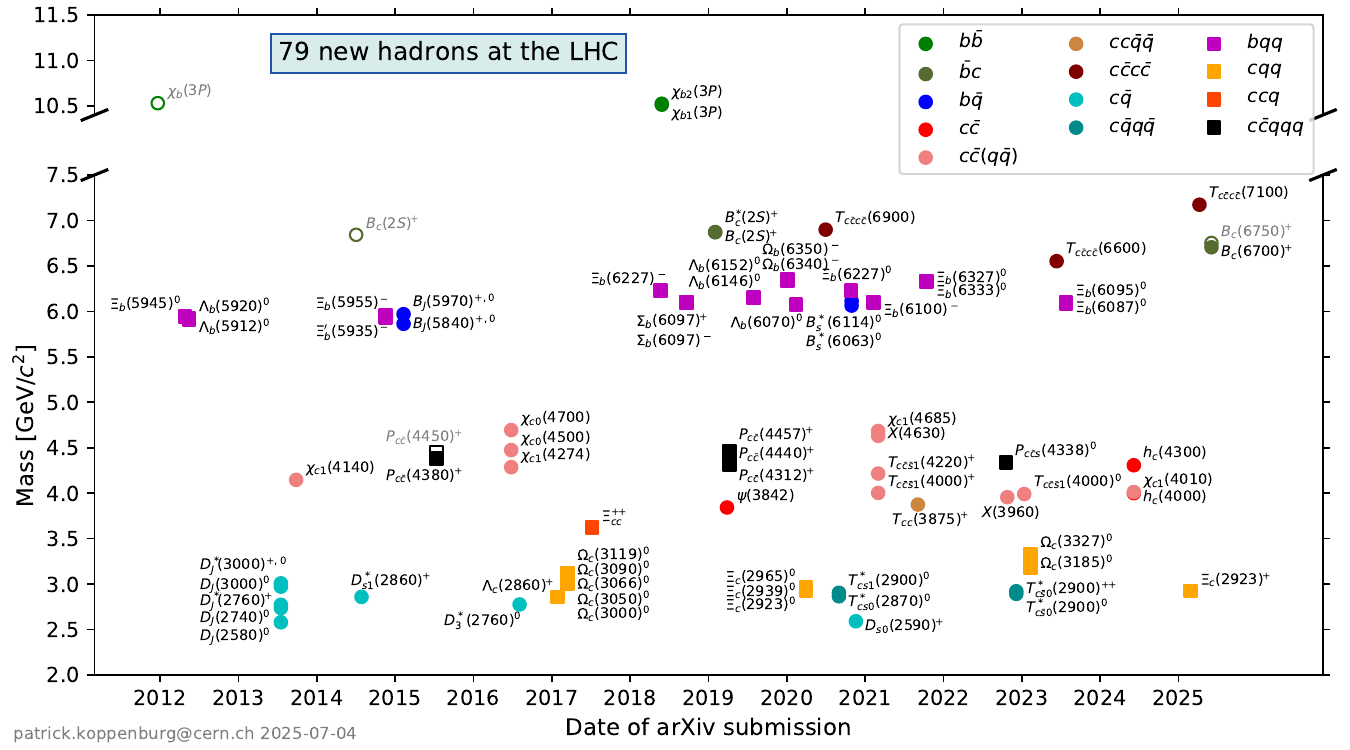}%
    \includegraphics[width=0.35\linewidth]{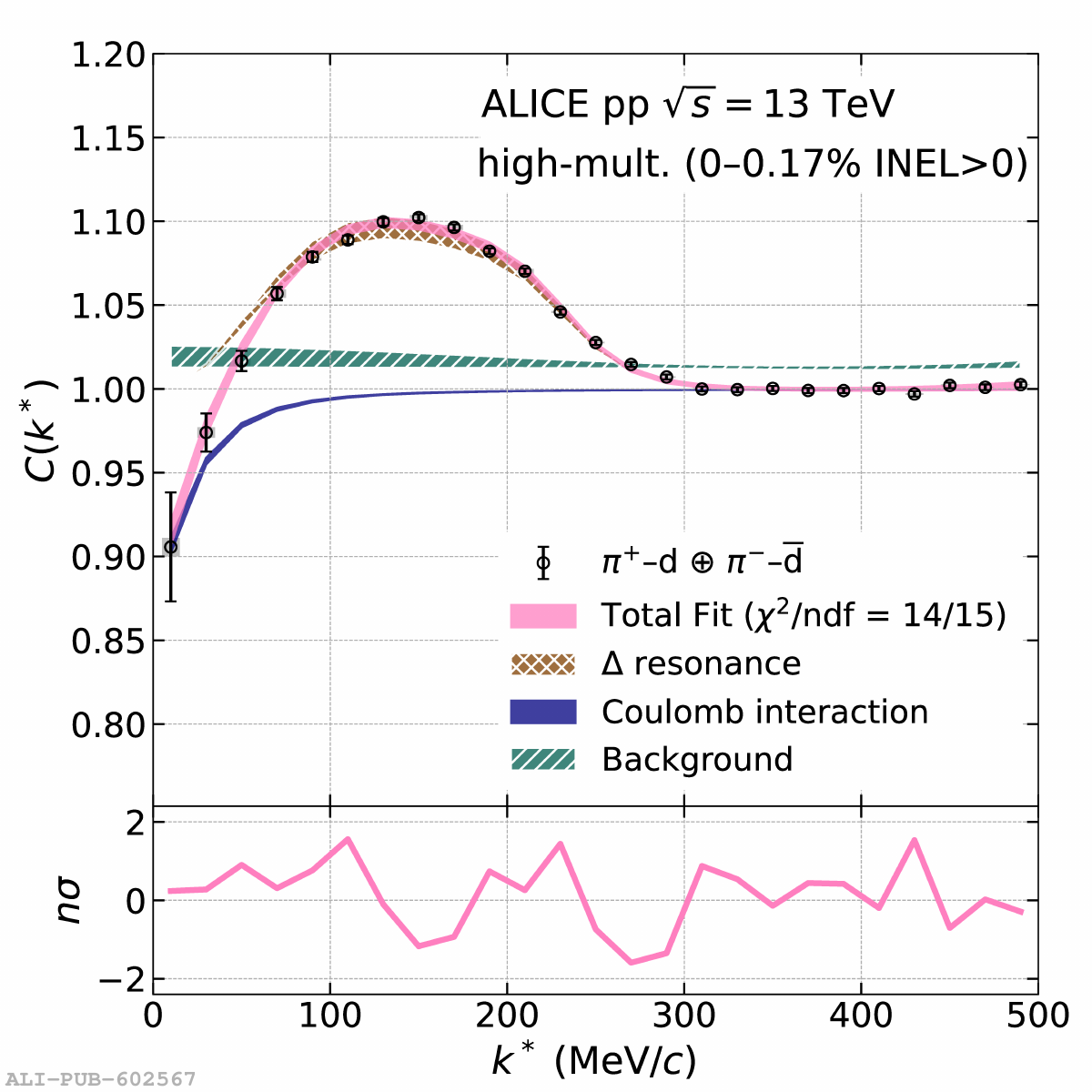}
    \caption{New bound states uncovered at the LHC~\cite{Koppenburg-particles} and ALICE nucleosynthesis studies~\cite{ALICE:2025byl}.}
    \label{fig:lhc-hadron-masses}
\end{figure}

The mystery of confinement probes the heart of the Yang-Mills mass gap and flavor problems.
Why is there structure? 
The nineteenth century saw mysterious structure in chemical line spectra, which ultimately triggered the quantum revolution. 
Today, the rich hadron spectrum poses analogous questions~\cite{Koppenburg-particles} (Figure~\ref{fig:lhc-hadron-masses} left). 
Why is there flavor structure behind these QCD spectral lines?
Do massive glueballs exist as expected from a QCD mass gap?
Is this unexplained structure a harbinger for new paradigms and deeper principles?

\begin{figure}
    \centering
    \includegraphics[width=0.5\linewidth]{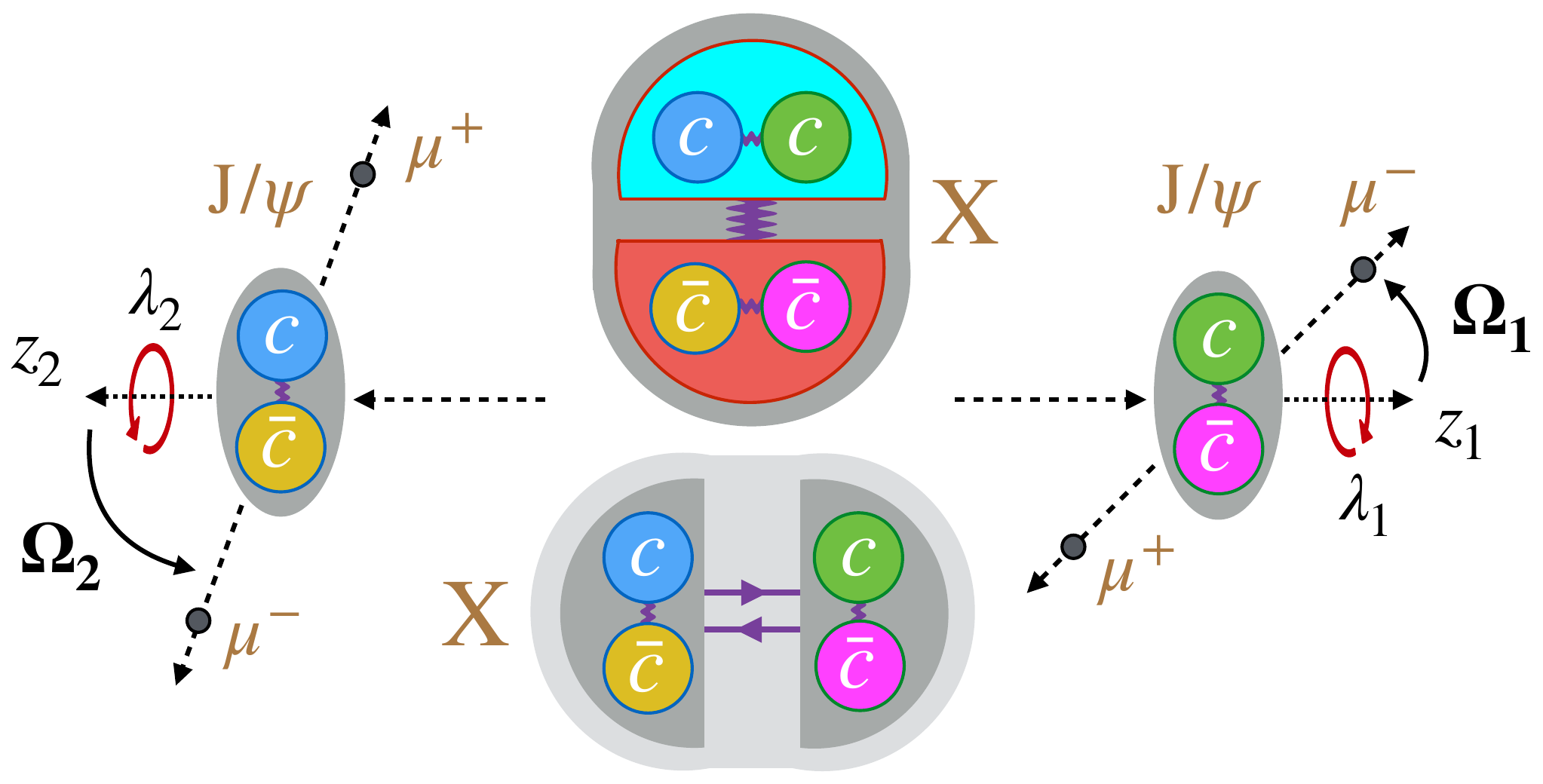}%
    \includegraphics[width=0.45\linewidth]{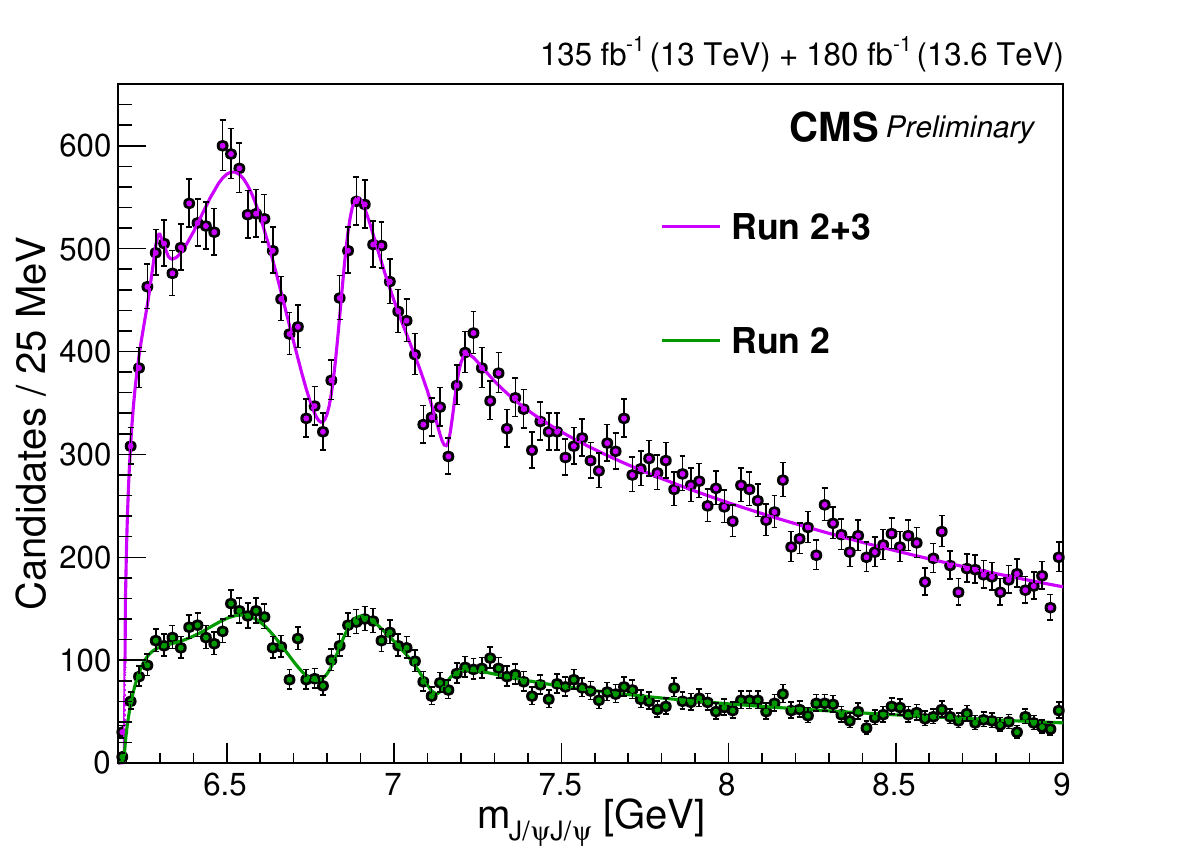}%
    \caption{Determination of quantum numbers for tetra-charm states~\cite{CMS:2025fpt}. }
    \label{fig:tetra-charm}
\end{figure}

\begin{figure}
    \centering
    \includegraphics[width=0.28\linewidth]{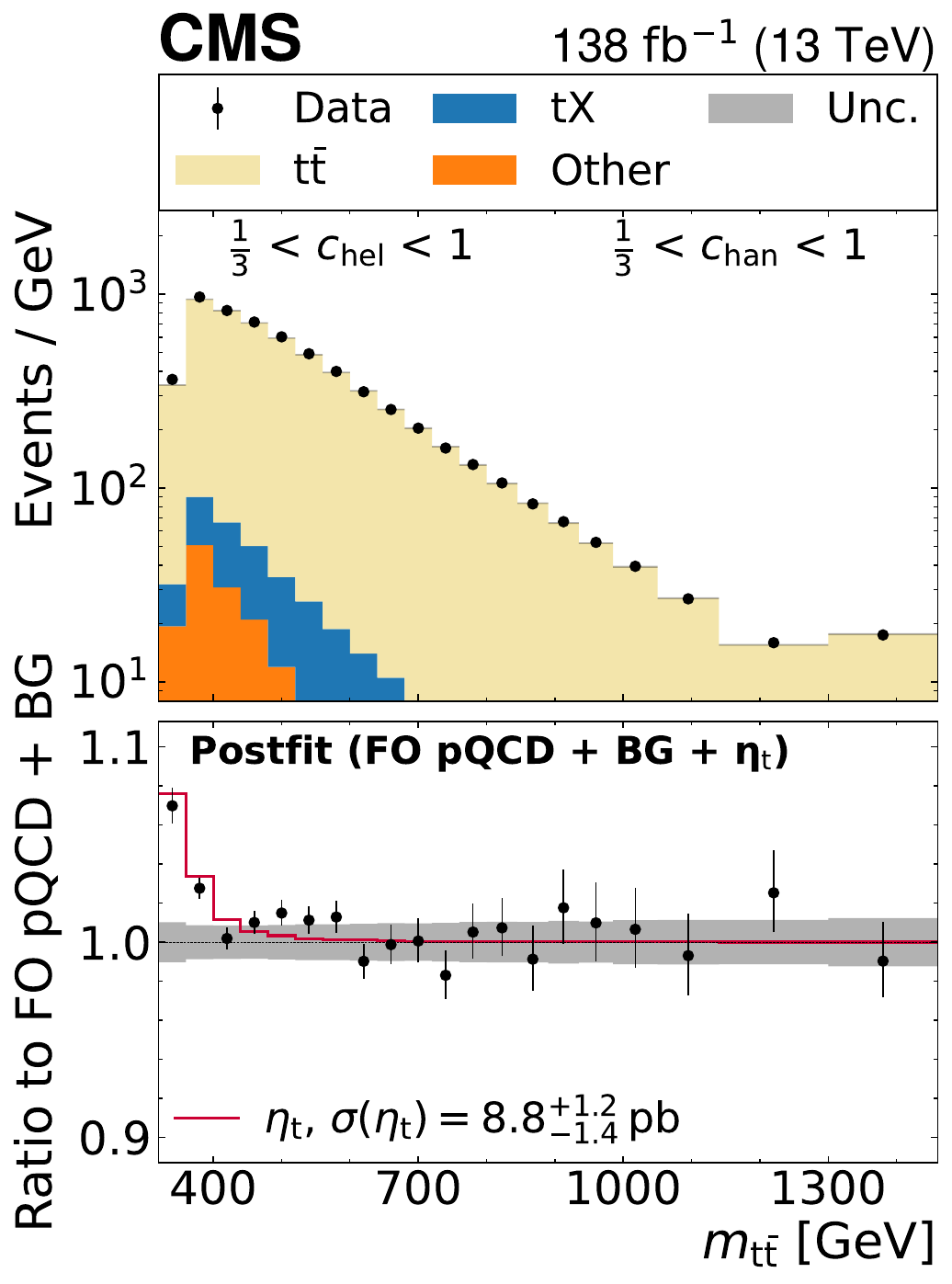}%
    \includegraphics[width=0.6\linewidth]{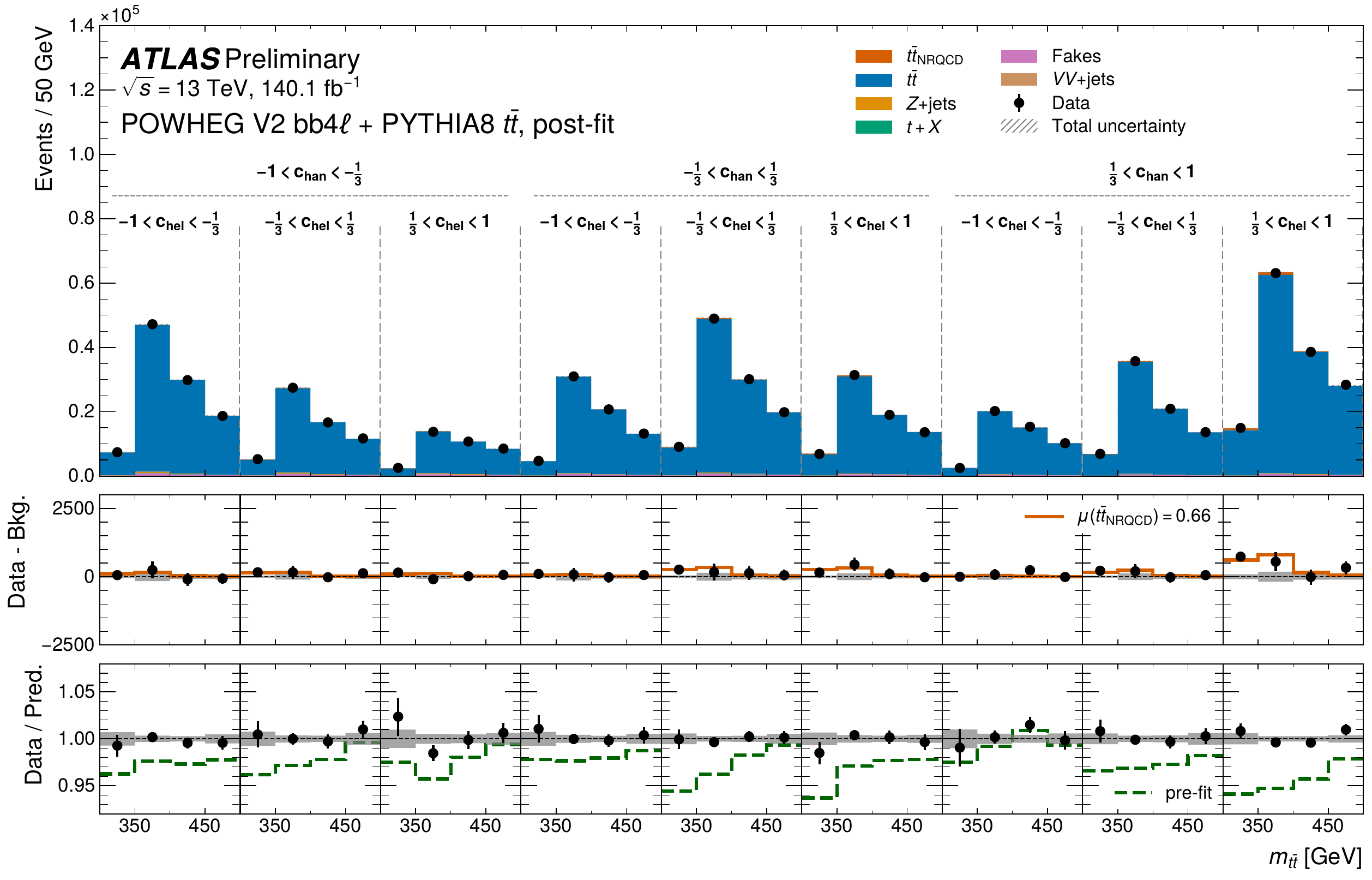}%
    \caption{Observed excess near the top-antitop threshold by CMS~\cite{CMS:2025kzt} and ATLAS~\cite{ATLAS-CONF-2025-008}. }
    \label{fig:toponium}
\end{figure}

Starting with light baryons, mysteries in nucleosynthesis endure.
ALICE released a new study to close long-standing gaps in nucleosynthesis important for cosmic-ray and dark matter science~\cite{ALICE:2025byl}.
How do nuclei with MeV binding energy form in conditions near pion scale temperatures?
They find model-independent evidence that 80\% of anti-deuterons form in nuclear fusion after $\Delta$ baryons decay.
Moving to heavier quarks, recent discoveries of all-charm tetraquarks need detailed characterization just like hydrogen in the nineteenth century.
CMS released a new study of all the spin parity and charge quantum numbers that constrains possible internal structure~\cite{CMS:2025fpt} (Figure~\ref{fig:tetra-charm}). 
Further studies may clarify interpretation as a bound state of four quarks or a molecule of two charmonium pairs.

Does the strong force also confine the heaviest fermion, the top quark?
No is the textbook answer, given its decay time is faster than confinement time $\tau_\text{decay} < \tau_\text{confinement}$.
But the uncertainty principle means there may be a non-zero probability for top quarks to momentarily bind in quasi-static regimes.
ATLAS recently confirmed~\cite{ATLAS-CONF-2025-008} the CMS excess~\cite{CMS:2025kzt} just below the top-antitop threshold (Figure~\ref{fig:toponium}) using a detailed spin correlation analysis to extract a signal consistent with spin-0 and odd parity. 
The precise interpretation of this excess  requires further study to clarify.

\section{\label{sec:cosmics}Extreme Cosmic-ray Puzzles}

\begin{figure}
    \centering
    \includegraphics[height=5cm]{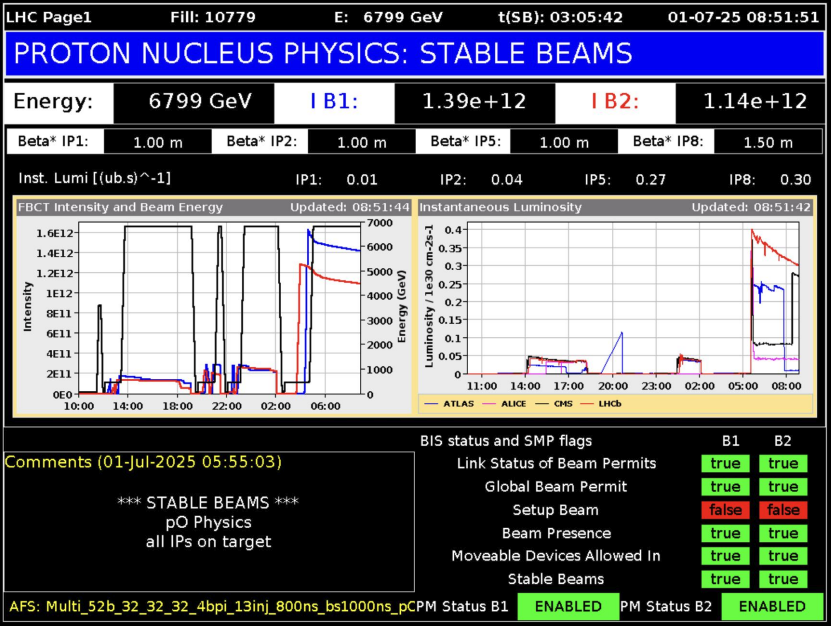}\;
    \includegraphics[height=5cm]{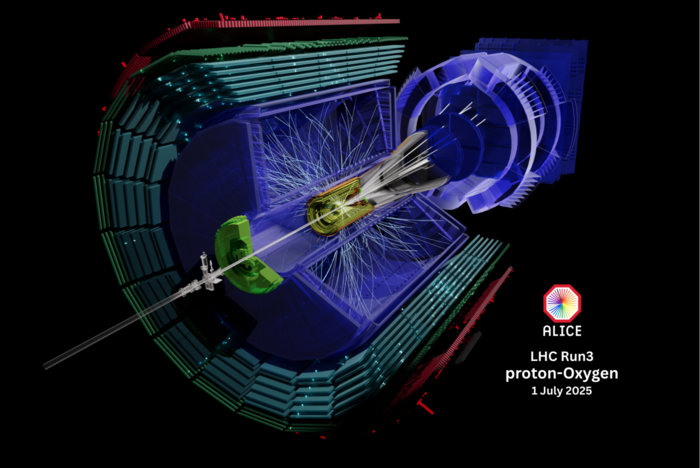}%
    \caption{World-first proton--oxygen collisions at the LHC in July 2025~\cite{alice-pO-displays}.}
    \label{fig:pO-collisions}
\end{figure}

An often overlooked but foundational connection between QCD and astrophysics is cosmic-ray science.
Precise QCD measurements at the LHC is critical for PeV astronomy.
Cosmic rays are the highest energy particles seen on Earth, but their origins remain enduring mysteries.
Where do they come from?
What are they made of?
How do they reach $10^{20}$ eV?
Air-shower observatories are the unique probe of these enigmatic ultra-high energy cosmic rays energies~\cite{PierreAuger:2017pzq,TelescopeArray:2023sbd}. 
However, poor air-shower modeling due to non-perturbative QCD is an obstructing systematic uncertainty.

\begin{figure}
    \centering
    \includegraphics[height=6cm]{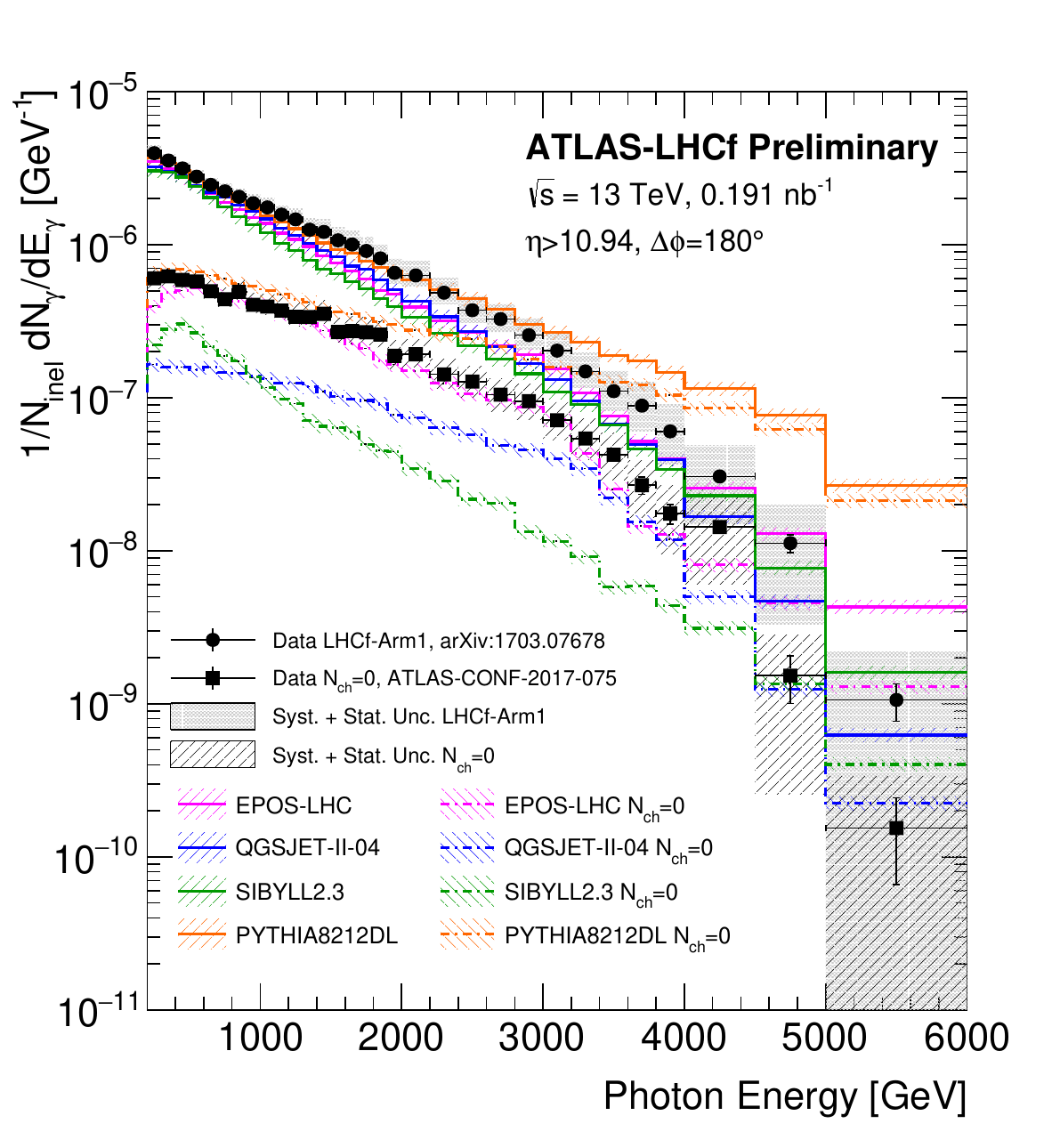}%
    \includegraphics[height=6cm]{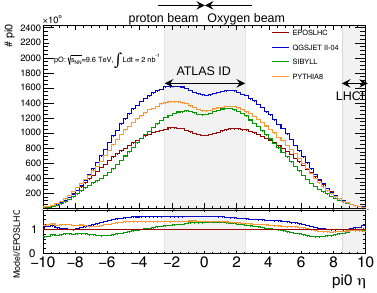}%
    \caption{Joint Run 2 ATLAS-LHCf analysis~\cite{ATLAS-CONF-2017-075} and simulated proton--oxygen pion spectra~\cite{leitgeb-lhcf-atlas}.}
    \label{fig:atlas-lhcf-joint-runs}
\end{figure}

\begin{figure}
    \centering
    \includegraphics[height=5cm]{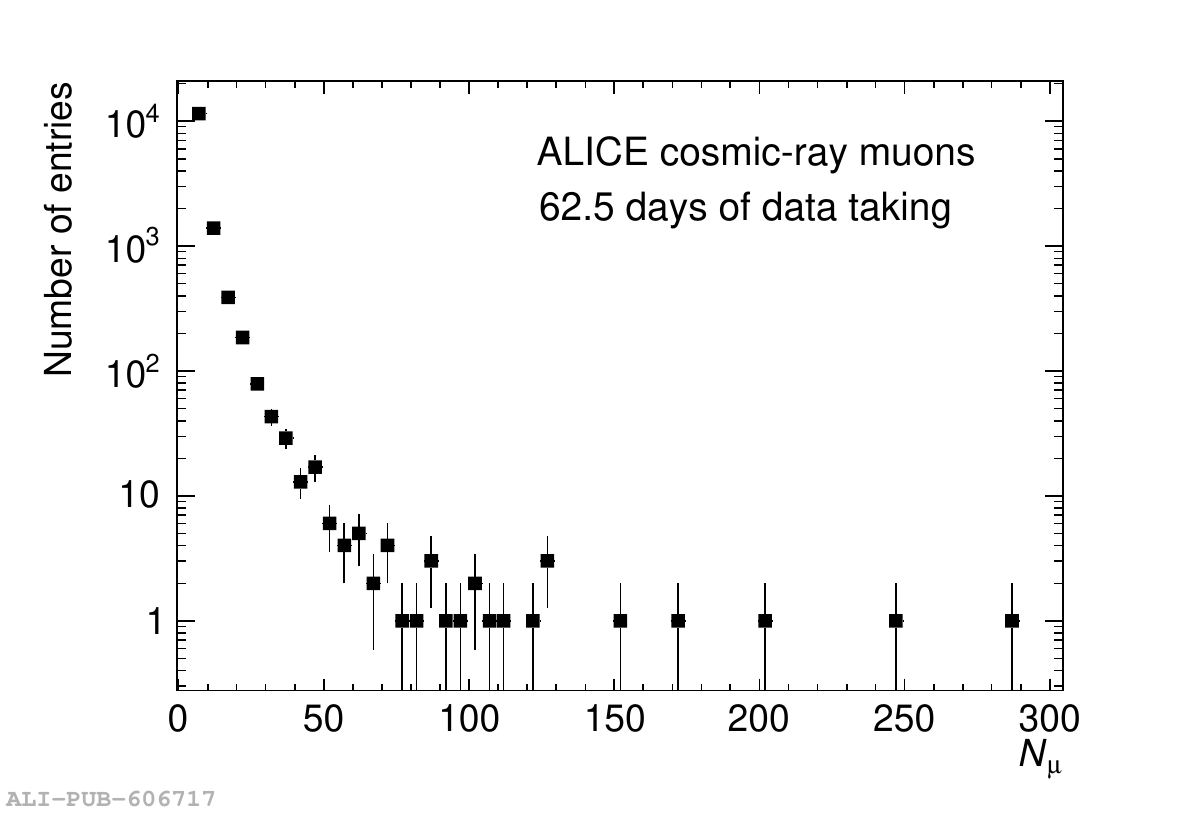}%
    \includegraphics[height=4.8cm]{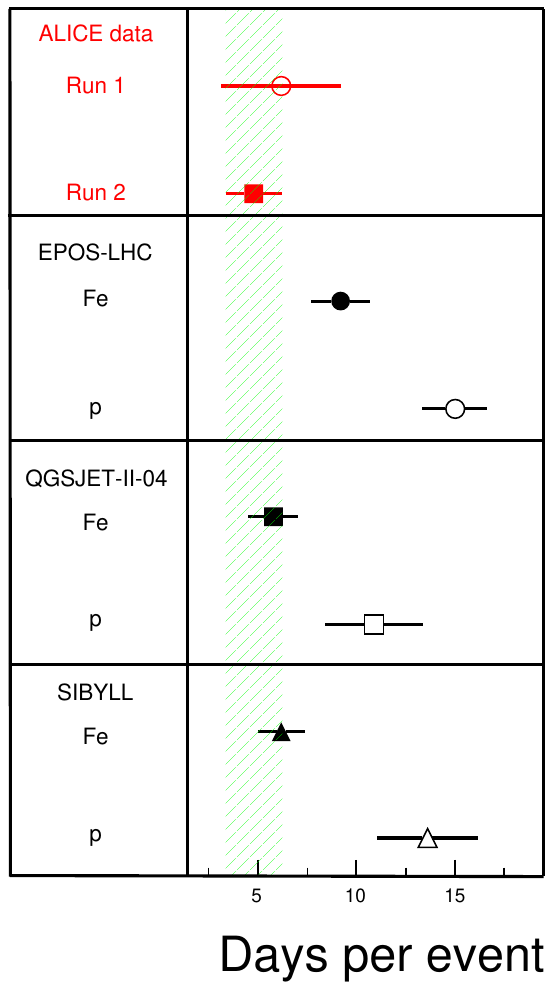}%
    \includegraphics[height=4.8cm]{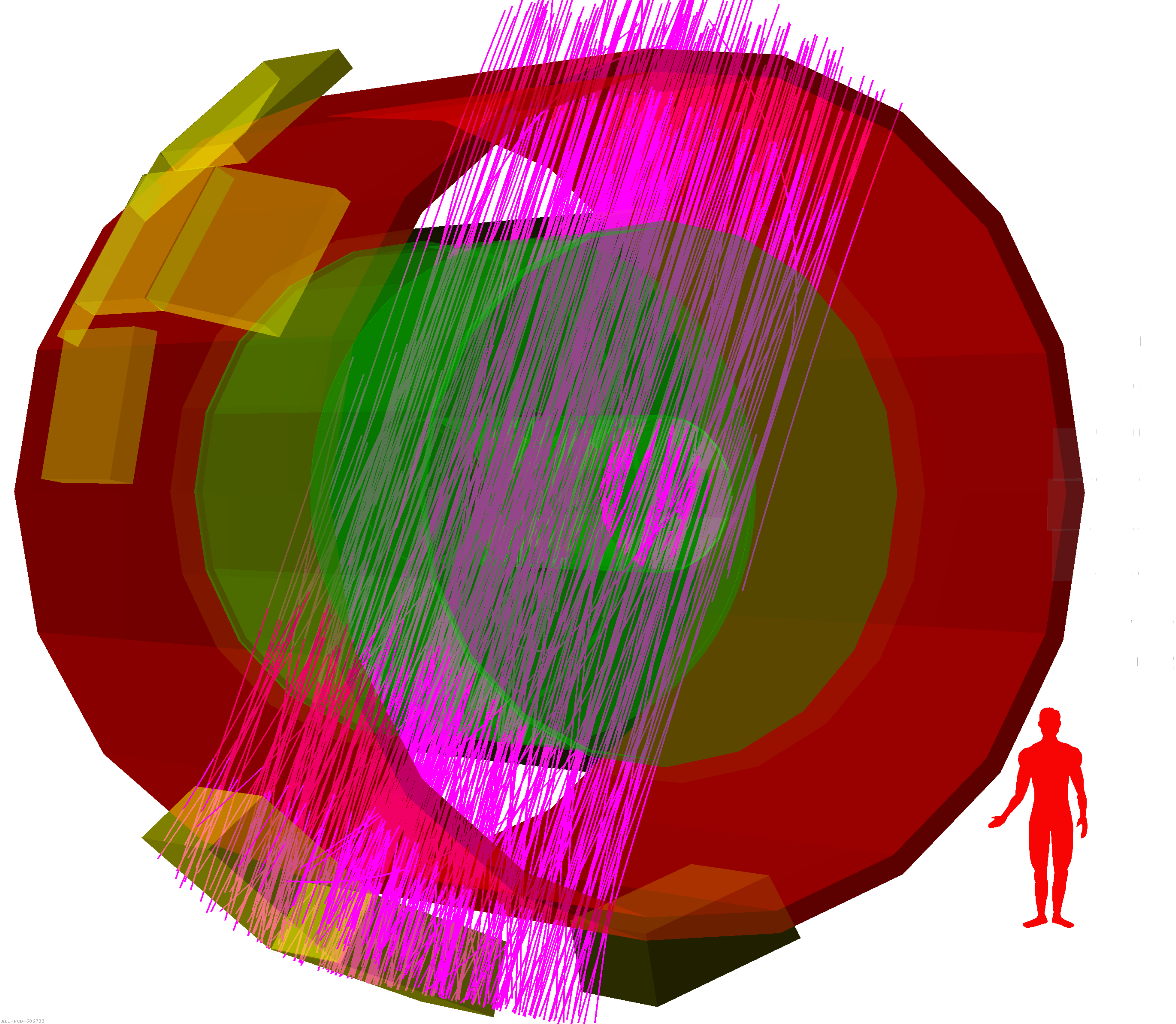}%
    \caption{ALICE measurements of multi-muon bundles in cosmic-ray events~\cite{ALICE:2024yqj}.}
    \label{fig:alice-cosmics}
\end{figure}

This motivates recreating PeV cosmic-ray showers in controlled laboratory conditions. 
In July 2025, the LHC pioneered world-first proton--oxygen collisions (Figure~\ref{fig:pO-collisions}), impossible at lepton colliders.
To capture the laboratory cosmic-ray shower,  special detectors are required.
Neutrals strike LHC Forward (LHCf) and Zero Degree Calorimeters (ZDC) inserted behind where the beam pipe forks, while the LHC dipoles sweep charged particles into Roman Pot spectrometers.
This enables joint inter-collaboration data taking.
In Run 2, ATLAS and LHCf piloted joint proton--proton data measuring photons out to 6~TeV~\cite{ATLAS-CONF-2017-075} (Figure~\ref{fig:atlas-lhcf-joint-runs} left).
These are the highest energy photons ever produced and measured in a laboratory.
In this regime, the data and models disagree not by tens of percent but by a shocking factor of a hundred.
Run 3 adopts the full suite of forward detectors to sharpen a 50\% spread in proton--oxygen models (Figure~\ref{fig:atlas-lhcf-joint-runs} right).
Analysis of these novel datasets with unconventional detectors are ongoing will improve PeV air-shower modeling.

Finally, even with the beam switched off, QCD science does not stop.
ALICE simply counts muons in 63 days of cosmics data to test QCD models tuned to LHC collision data~\cite{ALICE:2024yqj} (Figure~\ref{fig:alice-cosmics}).
The striking event display shows 287 cosmic-ray muons gracefully crossing the ALICE Time Projection Chamber (Figure~\ref{fig:alice-cosmics} right), with data favoring the  chemical composition cosmic-ray primary to have heavy components illustrated by predictions with iron.
ALICE is not just a collider detector but also a cosmic observatory, both a microscope and telescope.

\section{\label{sec:conclusion}Conclusions}

Recent LHC results epitomize remarkable advances in measurement science.
But to what end?
Why measure the next decimal point when theory predicts nothing?
History offers motivating lessons.
In 1928, the Dirac equation predicts the gyromagnetic factor of the electron is exactly two $g_e=2$, while the neutron discovered soon after in 1932 is exactly zero $g_n = 0$. 
Experimentalists could be criticized for pursuing measurements just to affirm zero. 
But in 1948, Kusch and Foley revealed groundbreaking nonzero deviations at per mille $g_e = 2.0023 \pm 0.0001$~\cite{Kusch:1948mvb} justifying one-loop Quantum Electrodynamics $\alpha_\text{EM}/\pi$~\cite{Schwinger:1948iu}.
This surprise revealed the vacuum is neither static nor empty as classically assumed, but a teeming sea of virtual particles embodying quantum fields.
Meanwhile, the neutron completely confounded expectations, being large and negative $g_n=-3.8$.
With hindsight, this was the first indirect evidence for quark confinement due to a new force: Quantum Chromodynamics. 
Measurement despite theory predicting nothing triggered profound paradigm shifts. 

Recent experimental progress render profound discoveries much likelier at higher precision with the ATLAS and CMS upgrade program at the High-Luminosity LHC~\cite{CERN-LHCC-2015-020,Aberle:2749422,PhaseII:ITkStripsTDR,Fomin:2025ykx}. 
The extraordinary breadth of these results underscore how the LHC is a transformative QCD laboratory illuminating the deepest quantum enigmas of the terascale.
Enduring mysteries remain as it is clear the strong force is not inevitable from first principles.
Experiment is needed to elucidate our place in the universe.

\emph{Acknowledgments}. JL is grateful to the Faculty of Arts and Sciences at New York University for research travel support and the University of Wisconsin--Madison for conference organization. 
Copyright 2025 CERN for the benefit of the ATLAS Collaboration. CC-BY-4.0 license.


\bibliographystyle{JHEP}
\bibliography{bibs/intro.bib,bibs/pheno.bib,bibs/software.bib,./bibs/exp.bib,./bibs/theory.bib,./bibs/upc.bib}

\appendix
\clearpage

\end{document}